\documentclass[iop]{emulateapj}

\begin{document}

\shorttitle{Heartbeat state in GRS~1915+105}

\title{A statistical analysis of the ``heartbeat" behaviour of GRS~1915+105}

\author{Shan-Shan Weng$^{1}$, Ting-Ting Wang$^{1}$, Jing-Ping Cai$^{1}$, Qi-Rong Yuan$^{1}$, Wei-Min Gu$^{2}$}

\affil{$^{1}$ Department of Physics and Institute of Theoretical Physics,
Nanjing Normal University, Nanjing 210023, China}

\affil{$^{2}$ Department of Astronomy, Xiamen University, Xiamen, Fujian
361005, China}

\email{wengss@njnu.edu.cn}
\begin{abstract}

GRS~1915+105 has been active for more than 26 years since it was discovered in
1992. There are hundreds of {\it RXTE} pointed observations on this source, and
the quasi-regular flares with a slow rise and a sharp decrease (i.e. the
``heartbeat" state) were recorded in more than 200 observations. The
connections among the disk/corona, jet, and the disk wind at the heartbeat
state have been extensively studied. In this work, we firstly perform a
statistical analysis of the light curves and the X-ray spectra to investigate
this peculiar state. We calculate the parameters for heartbeat cycles,
including the recurrence time, the maximum and the minimum count rate, the
flare amplitude, and the cumulative radiation for each cycle. The
recurrence time has a bimodal distribution ranging from $\sim 20$ to $\sim 200$
s. The minimum count rate increases with increasing recurrence time;
while the maximum count rate remains nearly constant around 2 Crab. Fitting the
averaged spectrum for each observation, we find the strong correlations among
the recurrence time, the apparent inner radius of the accretion disk (or
the color correction factor), and the (nonthermal) X-ray luminosity. We
suggest that the true inner edge of the accretion disk might always extend to the
marginally stable orbit, while the change in corona size should result in the
observed correlations.

\end{abstract}

\keywords{accretion, accretion disks --- black hole physics --- X-rays: binaries ---
X-rays: stars --- X-rays: individual (GRS~1915+105)}

\section{Introduction}

The prototype microquasar GRS~1915+105, consisting of a black hole (BH) and a
K-M \textsc{iii} companion star \citep{Greiner01}, was discovered by {\it
GRANAT}/WATCH in 1992 \citep{Castro92}. Since then, the X-ray monitoring data
reveal that it has remained bright during the last 26 years. The parallax
distance is estimated at $\sim 8.6$ kpc, and the dynamical BH mass is measured
as $\sim 12.4 M_{\odot}$ with an orbital period of 34 days \citep[][and
references therein]{Reid14}. GRS~1915+105 exhibits fantastic phenomena and thus
attracts a lot of attention to observations and accretion theories \citep[e.g.
][]{Fender04, Zdziarski05, McClintock06, Miller13}. Both the continuum
spectral fitting and the relativistic disk reflection measurement indicate a
rapidly spinning BH harbored in GRS~1915+105 \citep[e.g. ][]{Zhang97,
McClintock06, Middleton06, Blum09, Miller13, Reid14}. The powerful jet with
the apparent superluminal motion has been revealed \citep{Mirabel94}, and
whether the jets are driven by the BH spin energy is still under dispute
\citep{Fender10, Narayan12, Miller13}. The X-ray emissions show
dramatic variability and quasi-periodic oscillation (QPO) in different time
scales \citep[e.g. ][]{Morgan97, Belloni13, Zhang15}; therefore, GRS~1915+105
offers a remarkable laboratory for the investigation of accretion flows.

Analyzing a large set of {\it RXTE} observations, \cite{Belloni00} classified
12 different variability patterns of X-ray emission, based on the count rate
and the hardness ratio characteristics. Two additional variability classes were
proposed later by \cite{Klein02} and \cite{Hannikainen05}. Of these 14 classes
of X-ray variability, the most intriguing is the $\rho$ class, which displays
the quasi-periodic bright flares, being analogous to the ``heartbeat" behavior.
Therefore, the $\rho$ class is also referred to as the heartbeat state, and it
is only found in two BH X-ray binaries (XRBs), i.e. GRS~1915+105 and
IGR~J17091-3624 \citep[e.g. ][]{Altamirano11, Court17}. During the heartbeat
state, the X-ray luminosity approaches the Eddington luminosity of the source
at the peak of flares ($L_{\rm peak} \sim 10^{39}$ erg~s$^{-1}$). Thus, it provides an
opportunity for us to study the accretion flows being close to the Eddington
limit, which are extremely rudimentary.

Both the spectral and timing properties of the heartbeat state have been
extensively studied, and the connections among disk/corona, jet, and disk wind
have been investigated in many papers \citep[e.g. ][]{Neilsen11, Yan17, Yan18}.
But to date, there is a lack of statistical analysis on this peculiar state. In
this work, we investigate the light curves and the spectra of the heartbeat state
by using the full set of {\it RXTE} and search the correlations among the
different parameters (Section 2). We present our results in Section 3 and
explore the accretion theories at high luminosities close to the Eddington
limit with the obtained relationships in Section 4.

\section{Data Reduction}

The Rossi X-ray Timing Explorer Mission ({\it RXTE}, 1995 December -- 2012
January) carries three instruments, the Proportional Counter Array (PCA), the
High Energy X-ray Timing Experiment, and the All-Sky Monitor. The primary
scientific objective of {\it RXTE} is to determine the origin of rapid
variation from astronomical X-ray sources. During its 16 years of science
operations, {\it RXTE} visited GRS~1915+105 frequently ($\sim$ 1800 times). In
this paper, our study relies on two standard mode data from the PCA. The
Standard 1 mode data have a time resolution of 0.125 s allowing us to depict
the light-curve profile at the heartbeat state; however, they do not have
energy information for a spectral fitting
\footnote{\url{https://heasarc.gsfc.nasa.gov/docs/xte/recipes/stdprod\_guide.html}}.
Alternatively, the Standard 2 mode data have 129 energy channels covering the
full range of the PCA detectors, but the time resolution of 16 s is comparable
to the heartbeat cycle, that is not fine enough for the timing analysis. Thus,
we use the Standard 1 mode data for the timing study and the Standard 2 mode
data for the time-averaged spectral fitting.

\begin{figure}
\centering
\includegraphics[width=8cm]{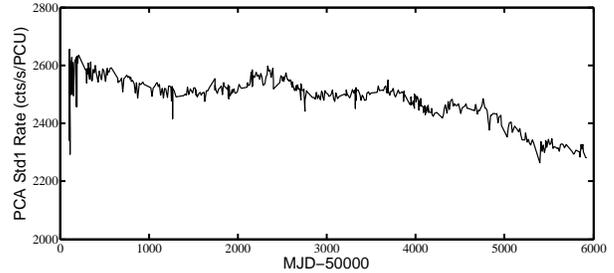}
\caption{{\it RXTE} mission long PCA light-curve from Crab. Because of the
degradation of detection efficiency, the count rate recorded in the Standard 1
mode data decreased from $\sim 2600$ cts/s/PCU to $\sim 2300$ cts/s/PCU.
\label{crab}}
\end{figure}

Different PCUs were active at different times because some of the PCUs suffered
breakdown and tripped off. In our temporal analysis, we adopt the standard
products of the Standard 1 mode data, which are normalized to 1 PCU. Since the
light curves of GRS~1915+105 could show chaotic variability at a short time
scale, we smooth them with a span of 2 seconds in order to reduce these
substructures. The light curves are further normalized to the
quasi-simultaneous count rate of the Crab Nebula
\footnote{\url{https://heasarc.gsfc.nasa.gov/docs/xte/recipes/mllc\_start.html}
to account for the degradation of detection efficiency (Figure \ref{crab})}. It
is worth noting that different flare classifications have been introduced by
different groups \citep[e.g. ][]{Yadav99, Belloni00, Massaro10}.
\cite{Massaro10} applied the Fourier and wavelet analysis to define the $\rho$
class, which could show regular and irregular variational patterns.
Alternatively, the irregular flares were classified as the $\kappa$ class, and
the regular flares were defined as the $\mu$ or $\rho$ classes in
\cite{Belloni00} according to the properties of the count rate and the X-ray
colors. In this paper, we adopt the classification in \cite{Belloni00}. That
is, the heartbeat state is referred to as the regular oscillation. We pick out
233 observations at the heartbeat state from all {\it RXTE} pointed data with
the visual inspection. The exposure time for these data ranges from 0.4 to 16.5
ks. For some observations with short exposure, it is difficult to distinguish
the $\rho$ from $\mu$ and $\kappa$ classes. Because of this, the number of
observations is slightly difference from those reported in \citet[][242
observations in their paper]{Neilsen12}.

The start time of oscillation ($T_{\rm start}$) is identified when the count
rate reaches the maxima in the interval of [$T_{\rm start} - \Delta T$, $T_{\rm
start} + \Delta T$], where $\Delta T$ is about 10\%-30\% of the recurrence
time. That is, we search the maxima by using a running box with a size of
$\sim$ 20\% of the recurrence time. Note that the derived $T_{\rm start}$ is
not affected by the size of the running box. It is very rare that the
recurrence time of some oscillation is almost twice as long as the neighbors.
In this case, the local maxima would be reported by the automated process, and
we would double check visually to screen out the local maxima. Once the $T_{\rm
start}$ is fixed, we introduce the structure parameters to characterize the
light curves as follows: the maximum (i.e. the peak, $C_{\rm max}$) and minimum
($C_{\rm min}$) count rates during each cycle; the recurrence time ($T_{\rm
rec}$) is defined as the time interval between two peaks; the amplitude of
flares is calculated as the ratio of $C_{\rm max}$ and $C_{\rm min}$ ($C_{\rm
max}/C_{\rm min}$); and the fluence is referred to as the cumulative X-ray
radiation during each cycle. Because the flare property is quite stable within
an individual observation, we calculate the mean value of these parameters for
every observation and assign the standard deviations to the errors of
parameters.

The exposure time of an individual pointing is much longer than $T_{\rm rec}$,
and one observation contains at least three heartbeat cycles. Therefore, the
spectrum from a full observation can be treated as the phase averaged spectrum
for the following analysis. We extract the X-ray spectrum with the Standard 2
mode data from the top layer of PCU2, which operated during all the
observations. The data are filtered with standard criteria. The background
files are produced with the bright source background model by using the task
\texttt{pcabackest}, and the response file is created with the PCA response
generator PCARMF v11.7. A systematic error of 0.5\% as recommended by the PCA
team is added in the spectral modeling. Following the work in \cite{Neilsen11},
we fit the deadtime corrected spectrum in the 3--30 keV with the empirical
model consisting of a multicolor disk, a power-law component plus a high-energy
cutoff \citep[{\it tbabs*highecut*(diskbb+powerlaw)} in XSPEC,][]{Arnaud96}.
The neutral hydrogen column density is fixed to $5\times10^{22}$ cm$^{-2}$
\citep{Lee02, Zoghbi16}. The yielded reduced $\chi^{2}$ centered around $\sim
0.74$ with a minima of 0.39 and a maxima of 1.27. No correlation between the
reduced $\chi^{2}$ and other spectral parameters is found. The unabsorbed
fluxes in 3--30~keV are estimated with the convolution model {\it cflux}, and
the errors for all spectral parameters are calculated in 68.3\% confidence
level.

\section{Results}

The goal of our work is to search the correlation between the heartbeat
parameters and to achieve the knowledge on the accretion at high luminosity
close to the Eddington limit.

\subsection{Light-curve analysis}
\begin{figure*}
\centering
\includegraphics[width=5.5cm]{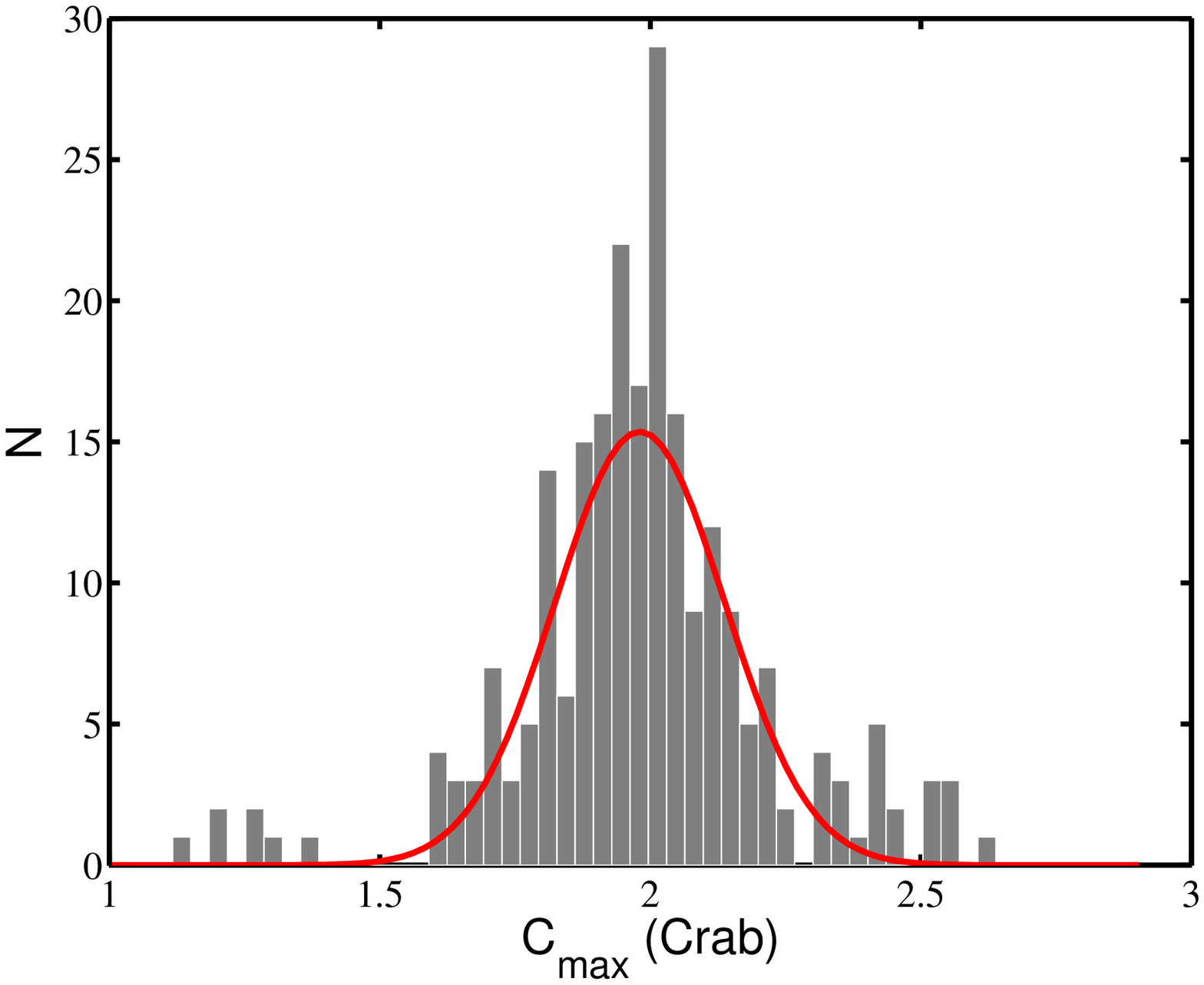}
\includegraphics[width=5.5cm]{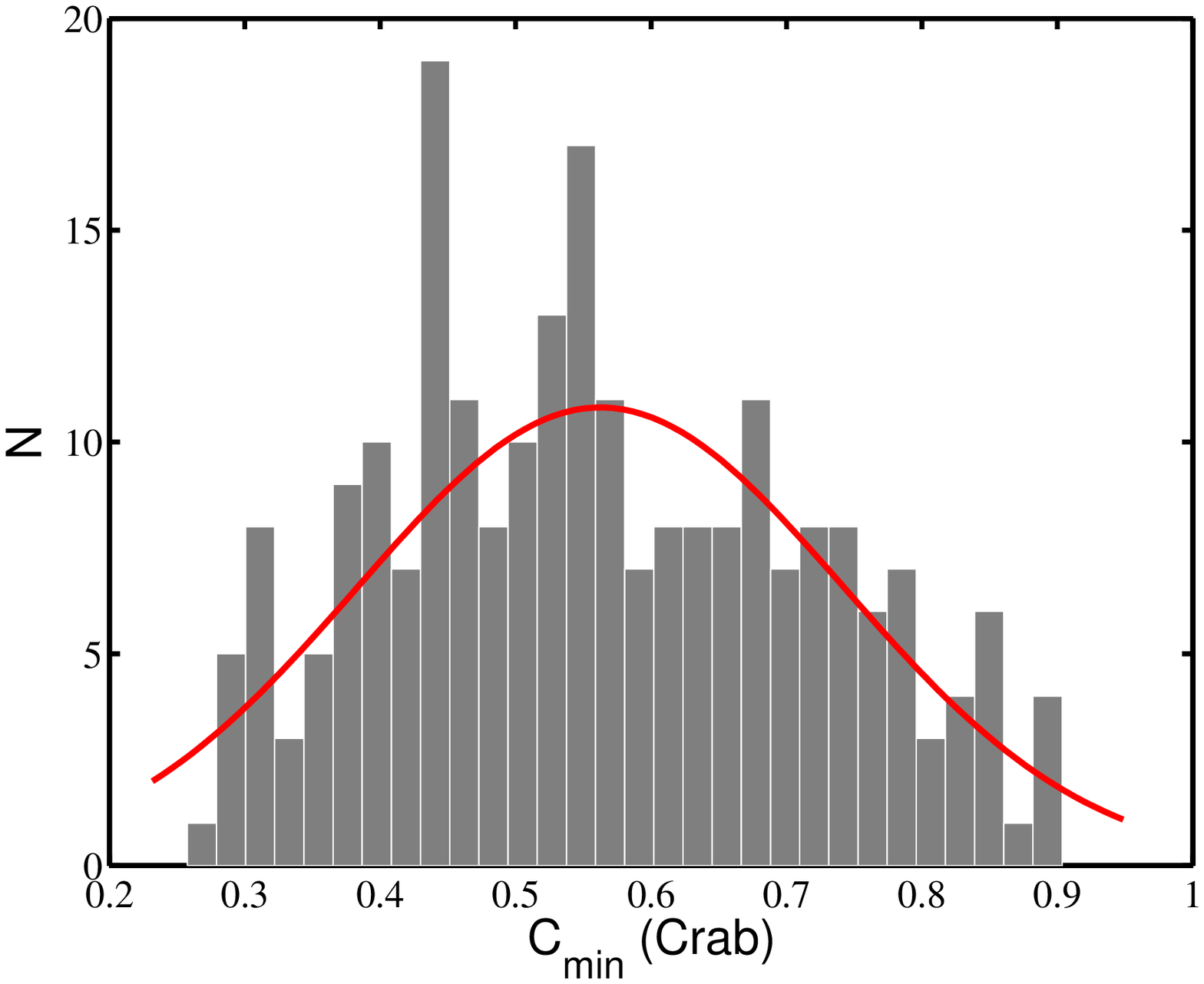}
\includegraphics[width=5.5cm]{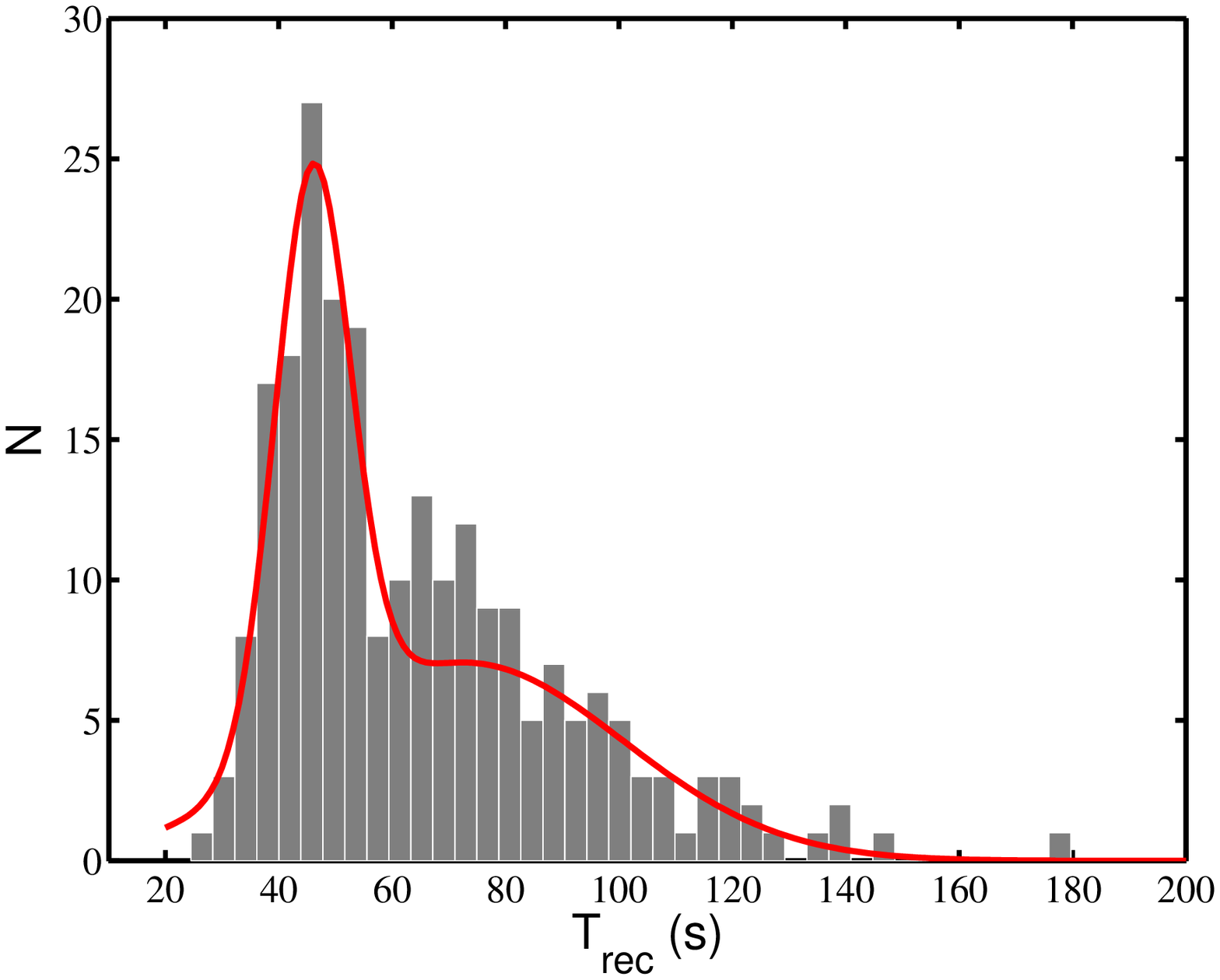}
\includegraphics[width=5.5cm]{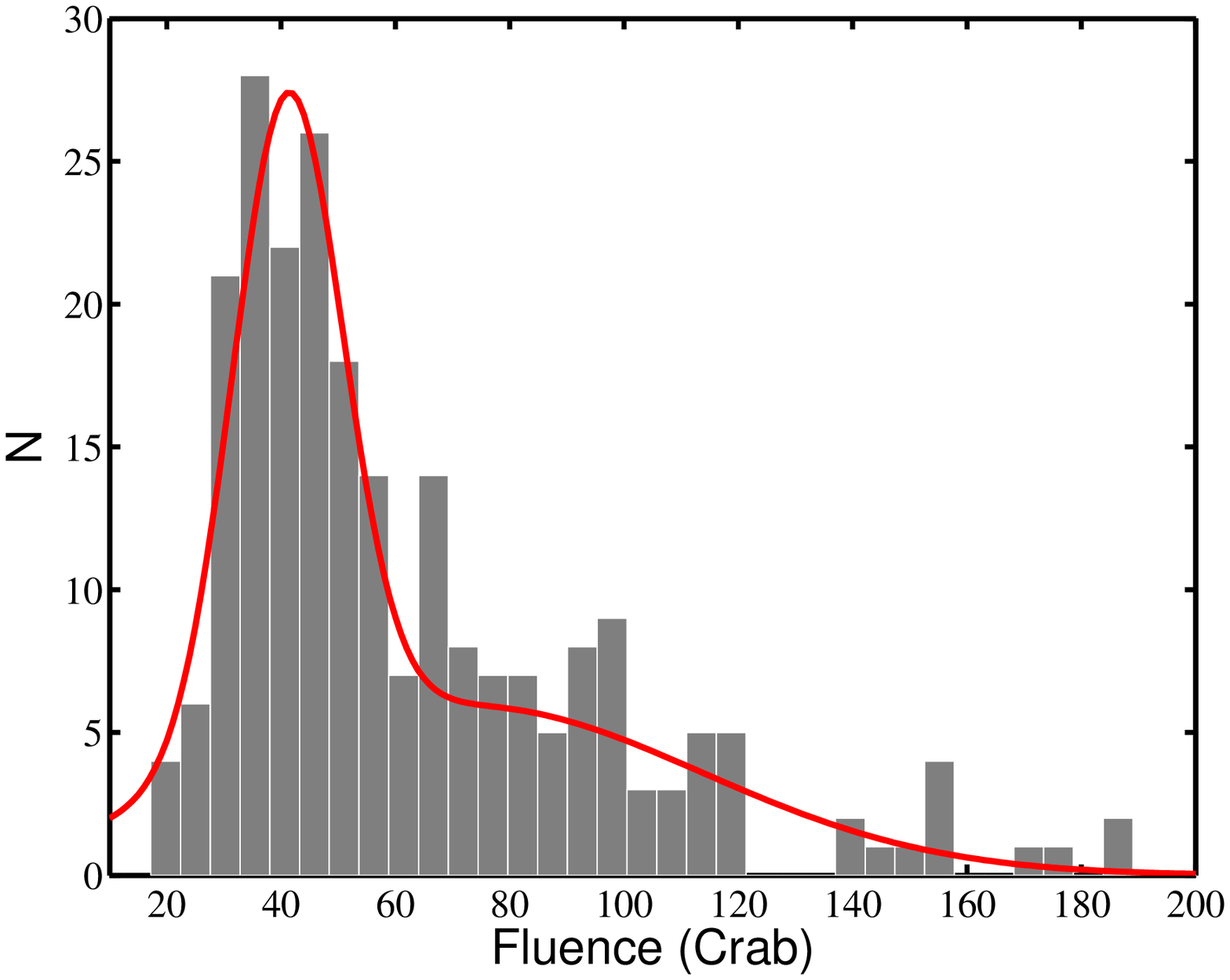}
\includegraphics[width=5.5cm]{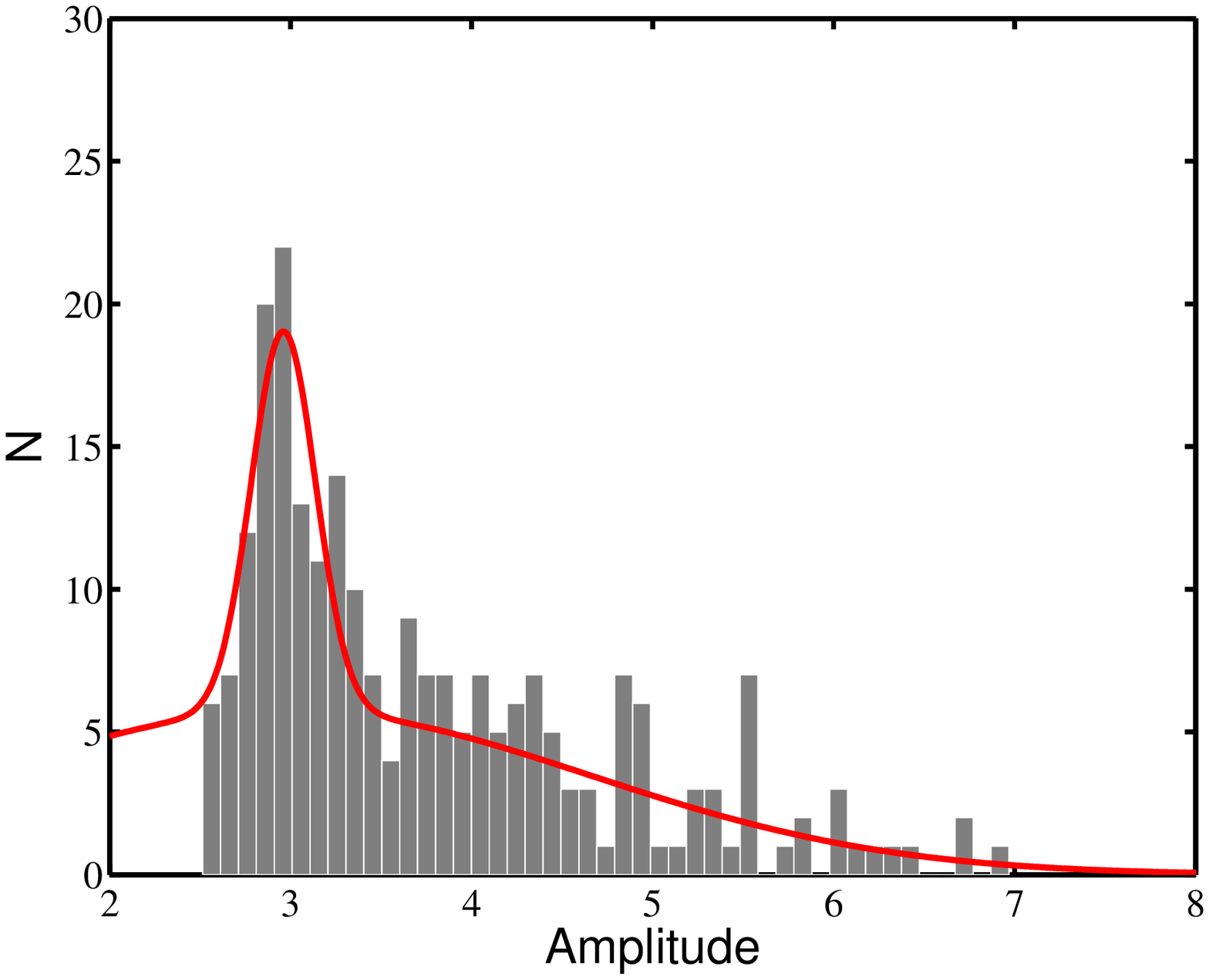}
\caption{Distributions of the light-curve structure parameters. We also apply
either the single-Gaussian or double-Gaussian fitting to these distributions
(red lines, Table \ref{profile}). \label{hist}}
\end{figure*}

\begin{figure*}
\centering
\includegraphics[width=6cm]{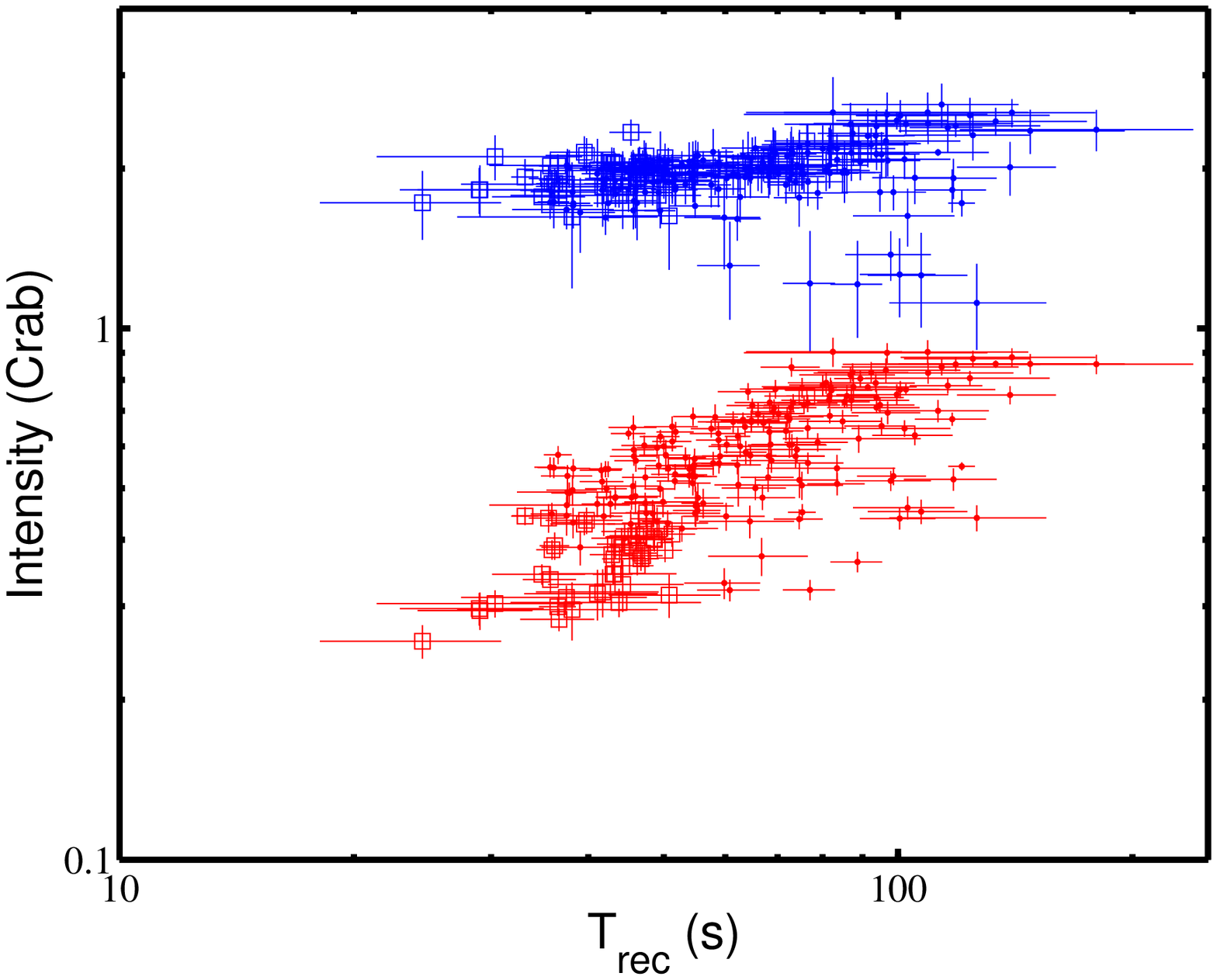}
\includegraphics[width=6cm]{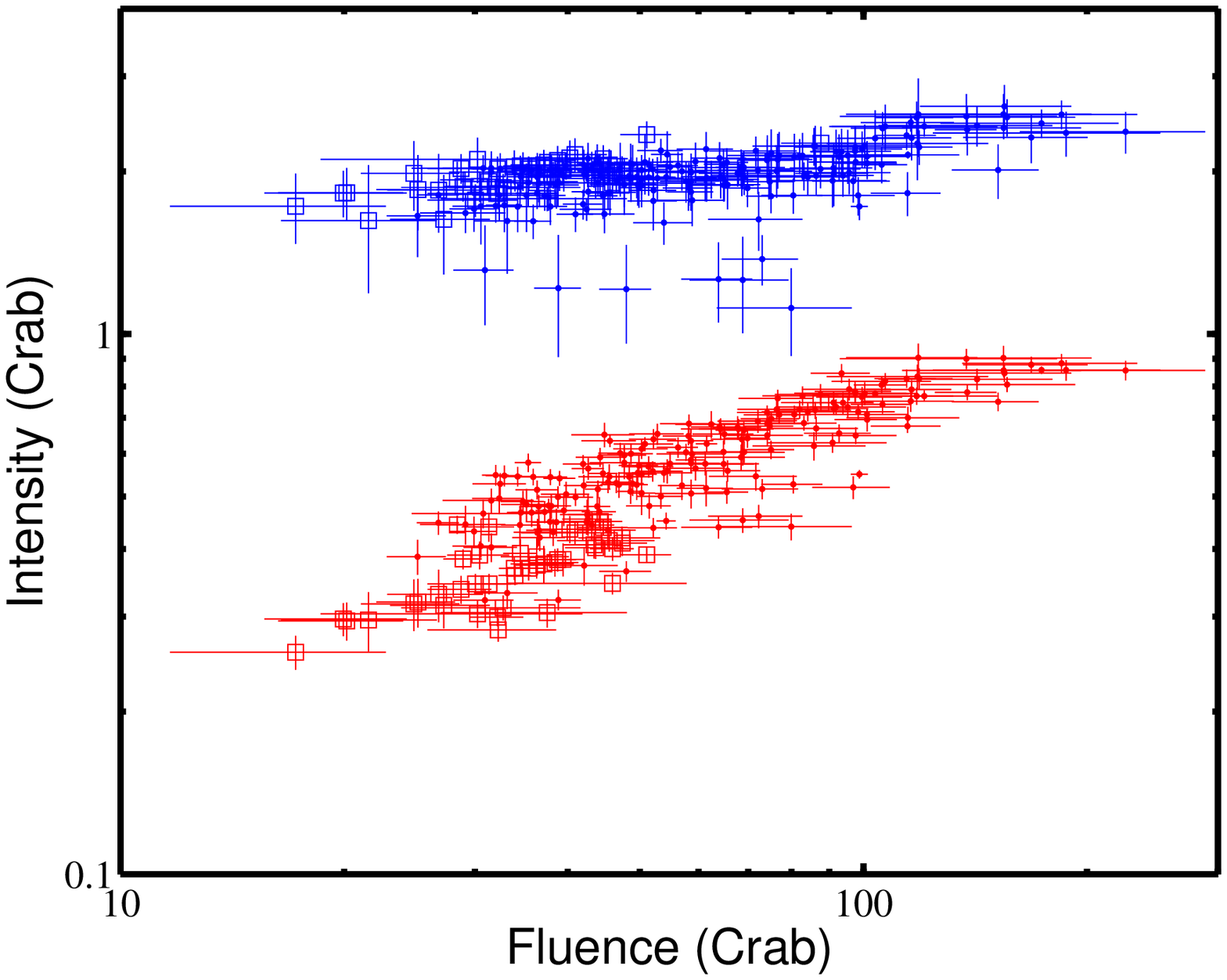}
\includegraphics[width=6cm]{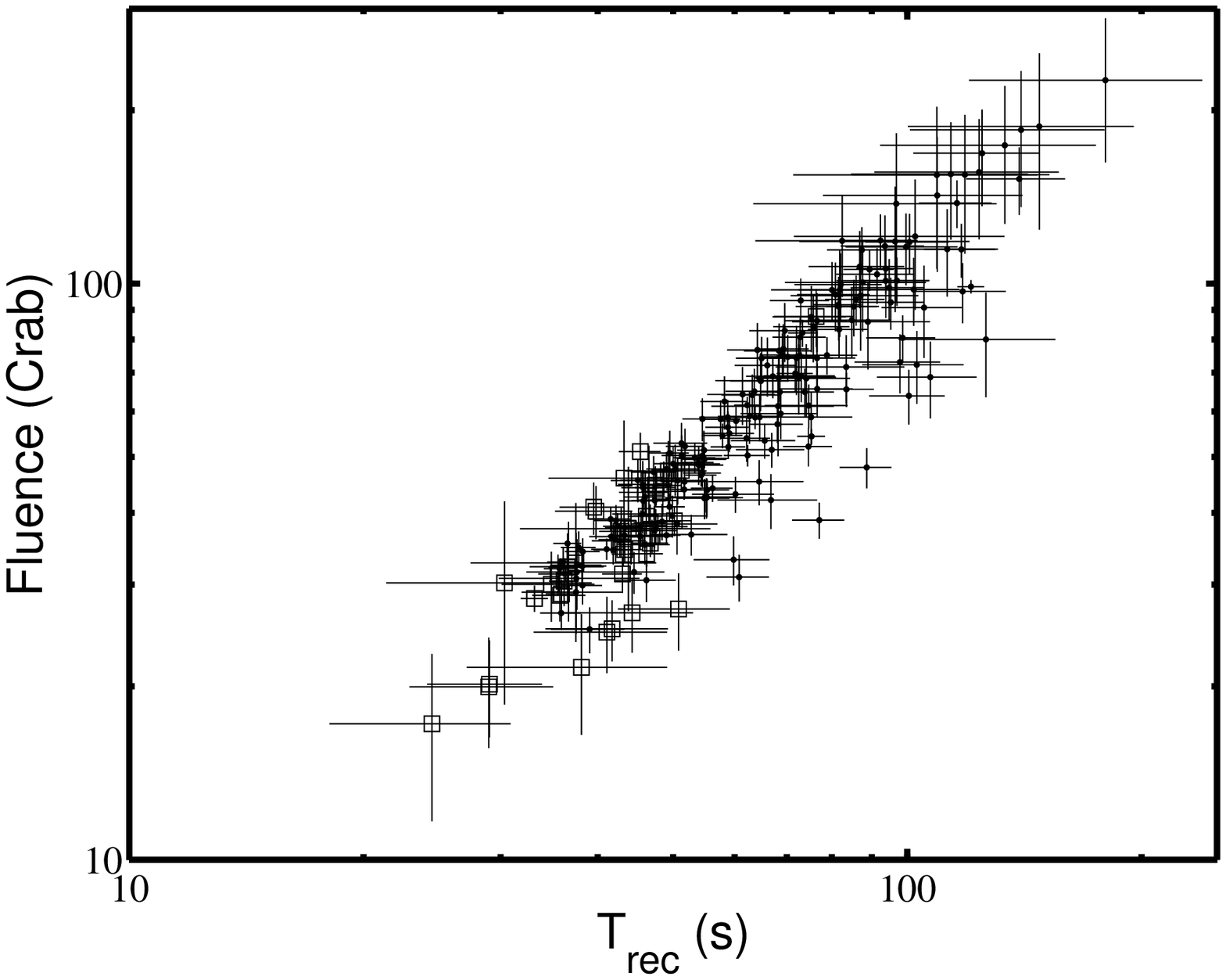}
\includegraphics[width=6cm]{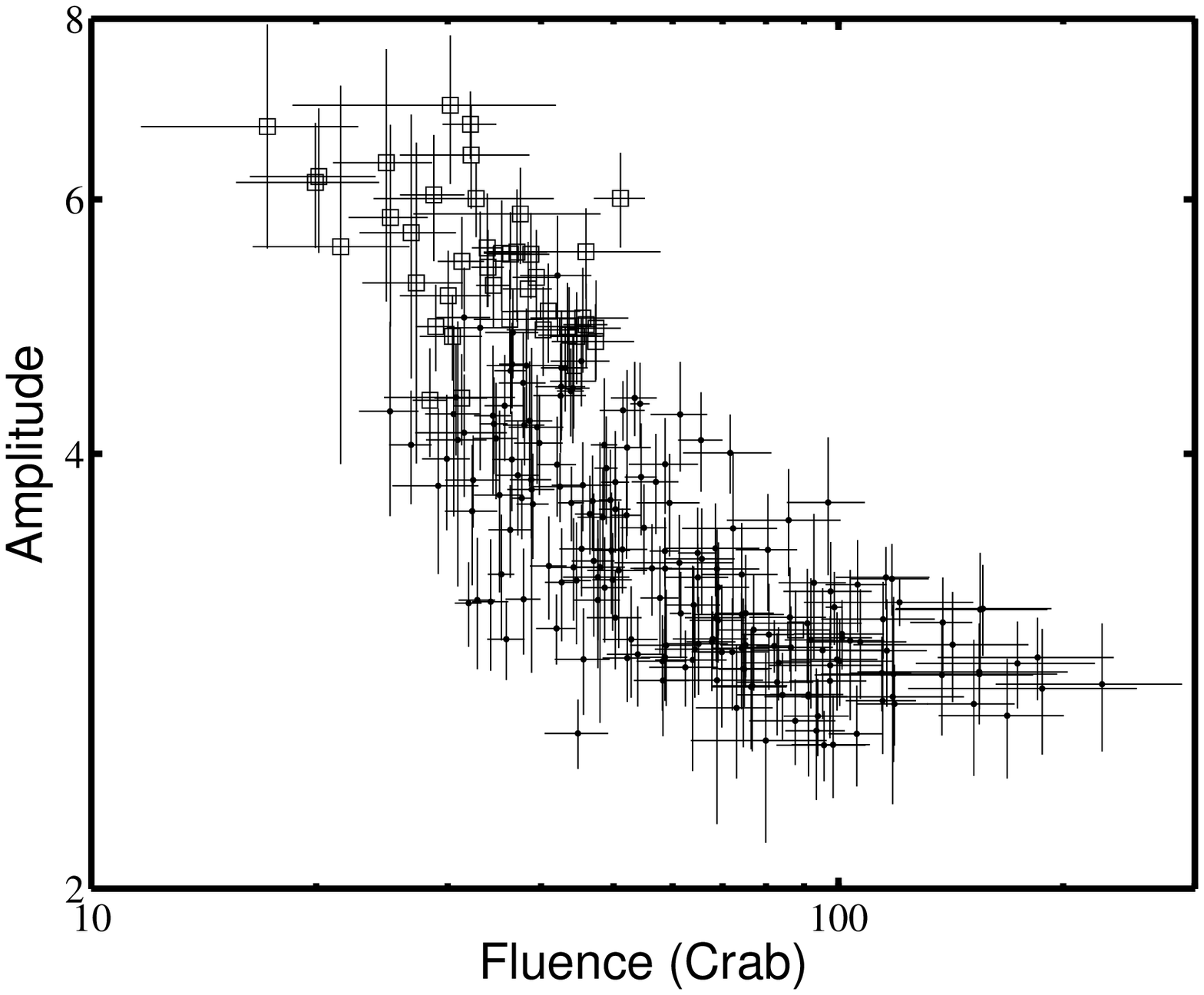}
\includegraphics[width=6cm]{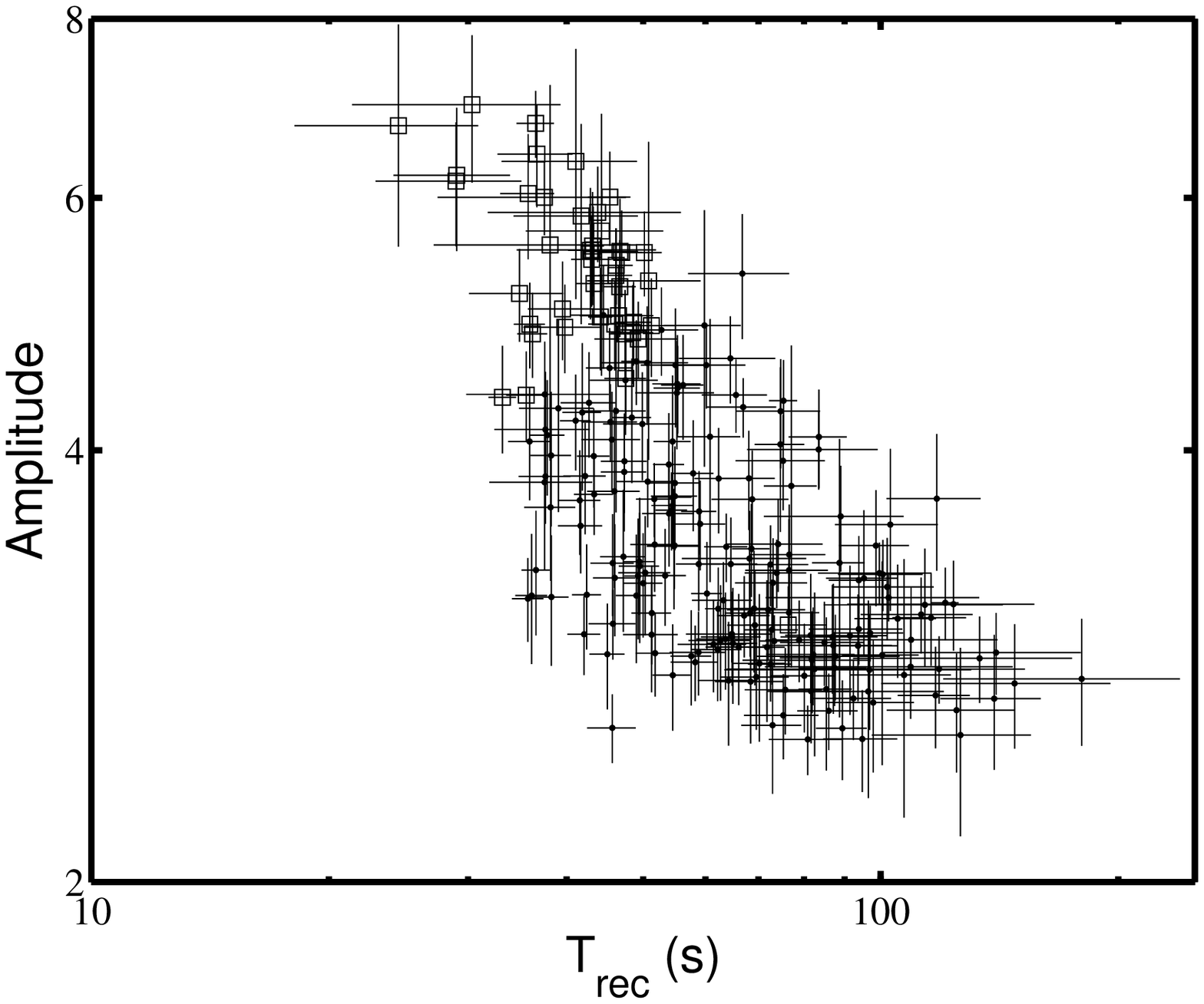}
\includegraphics[width=6cm]{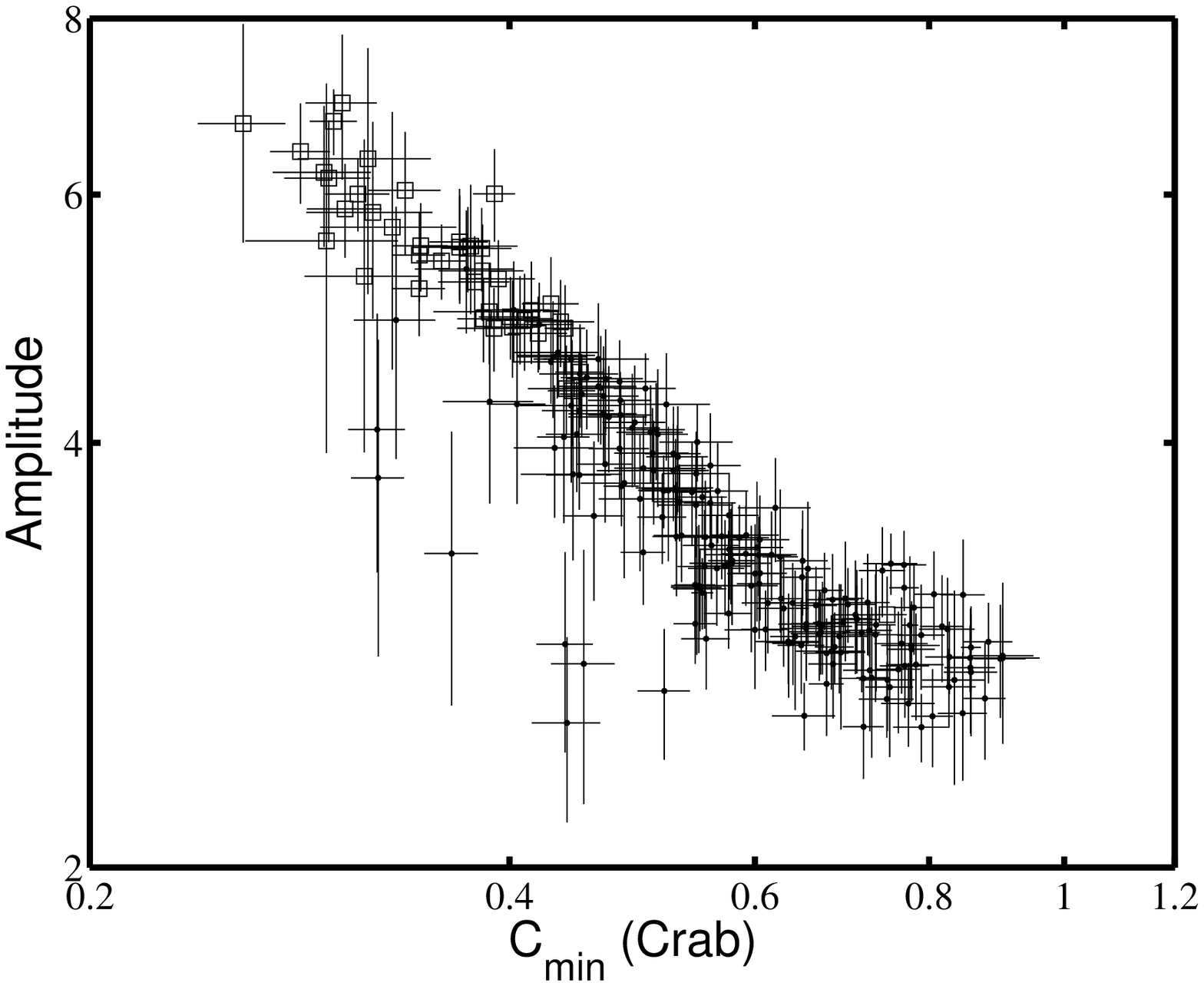}
\caption{Correlations between the structure parameters of light curves. In the
top two panels, the red and blue points correspond to $C_{\rm min}$ and $C_{\rm
max}$, respectively. In this work, open squares and filled cycles mark
the data with the apparent inner radius being smaller and larger than 70 km
(Figure \ref{spec_corr}), respectively. \label{lc_corr}}
\end{figure*}

The distributions of the time-varying parameters are plotted in Figure
\ref{hist}. The recurrence time ranges from $\sim 20$ to $\sim 200$ s, and the
peak count rates cluster around 2 Crab. The $C_{\rm min}$ and $C_{\rm max}$
distributions can be fitted by a single-Gaussian function, $f(x) =
a_{1}~e^{-(\frac{x-b_{1}}{c_{1}})^{2}}$, where $a_{1}$, $b_{1}$, and $c_{1}$
are fitting parameters. Alternatively, the double-Gaussian function ($f(x) =
a_{1}~e^{-(\frac{x-b_{1}}{c_{1}})^{2}} + a_{2}~
e^{-(\frac{x-b_{2}}{c_{2}})^{2}}$) provides better fittings to the
distributions of $T_{\rm rec}$, $Amplitude$, and $Fluence$ according to the
F-test estimation (with a significance level above 99.9\%). The fitting results
are shown in Table \ref{profile} and Figure \ref{hist}. The bimodal
distributions of $T_{\rm rec}$ and $Fluence$ are responsible for the two
branches having a small apparent inner disk radius ($R_{\rm in} \leq 70$ km)
and a large one ($R_{\rm in} > 70$ km, Figure \ref{spec_corr}, Section 3.2),
respectively. The branch with smaller $R_{\rm in}$ typically has smaller
$C_{\rm min}$, $Fluence$, and shorter $T_{\rm rec}$ (see also Figure
\ref{lc_corr}), and vice versa.

\begin{deluxetable*}{lcccccc}
\tabletypesize{\tiny} \tablewidth{0pt} \tablecaption{Best-fit  parameters for
the distributions of the light-curve structure parameters}
\tablehead{\colhead{}  & \colhead{$a_{1}$} & \colhead{$b_{1}$} &
\colhead{$c_{1}$} & \colhead{$a_{2}$} & \colhead{$b_{2}$} & \colhead{$c_{2}$}}
\startdata \hline
$C_{\rm max}$ & $15.4\pm2.2$  & $1.98\pm0.01$ & $0.22\pm0.02$  & --  & -- & --   \\
$C_{\rm min}$ & $10.8\pm1.2$  & $0.56\pm0.01$ & $0.25\pm0.02$  & --  & -- & --  \\
\hline
$T_{\rm rec}$ & $20.4\pm2.3$    & $45.9\pm0.8$  & $9.4\pm1.4$  & $7.1\pm0.8$  & $72.8\pm4.9$ & $39.4\pm4.2$  \\
$Fluence$     & $23.0\pm2.5$    & $41.1\pm1.5$  & $13.6\pm3.1$ & $5.9\pm1.4$  & $72.5\pm19.1$ & $58.3\pm15.3$\\
$Amplitude$   & $13.3\pm2.0$    & $2.96\pm0.04$ & $0.25\pm0.07$& $5.7\pm1.8$  &
$2.98\pm1.15$  & $2.4\pm0.8$
\enddata
\tablecomments{The distributions of $C_{\rm max}$ and $C_{\rm min}$ are fitted
by a single-Gaussian function. On the other hand, those of $T_{\rm rec}$,
$Amplitude$, and $Fluence$ are fitted by a double-Gaussian function.
\label{profile}}
\end{deluxetable*}

\cite{Mineo16} investigated two long {\it Beppo}SAX pointings, and suggested
that the minimum count rate was proportional to the one-third power of the
recurrence time \citep[see also ][]{Massaro10}. Here, we calculate the
Spearman's rank correlation coefficient of $\rho/P = 0.70/0$ between $C_{\rm
min}$ and $T_{\rm rec}$, confirming their positive correlation. When fitted
with a power-law function ($C_{\rm min} = a~T_{\rm rec}^{n}$), we obtain an
exponent of $n = 0.49\pm0.04$ and $a = 0.076\pm0.010$. Since the maximum count
rate remains nearly constant, the amplitude ($C_{\rm max}/C_{\rm min}$) is
anticorrelated with the $C_{\rm min}$ (Figure \ref{lc_corr}), which agrees with
the results reported in \cite{Naik02}. The fluence is correlated with $T_{\rm
rec}$ and is anticorrelated with the amplitude. Here, we also try to define the
heartbeat cycle as the time between the minimum count rate. Because all of the
structure parameters do not change much within an individual observation, the
results derived in this way remain unchanged.

\begin{figure*}
\centering
\includegraphics[width=12cm]{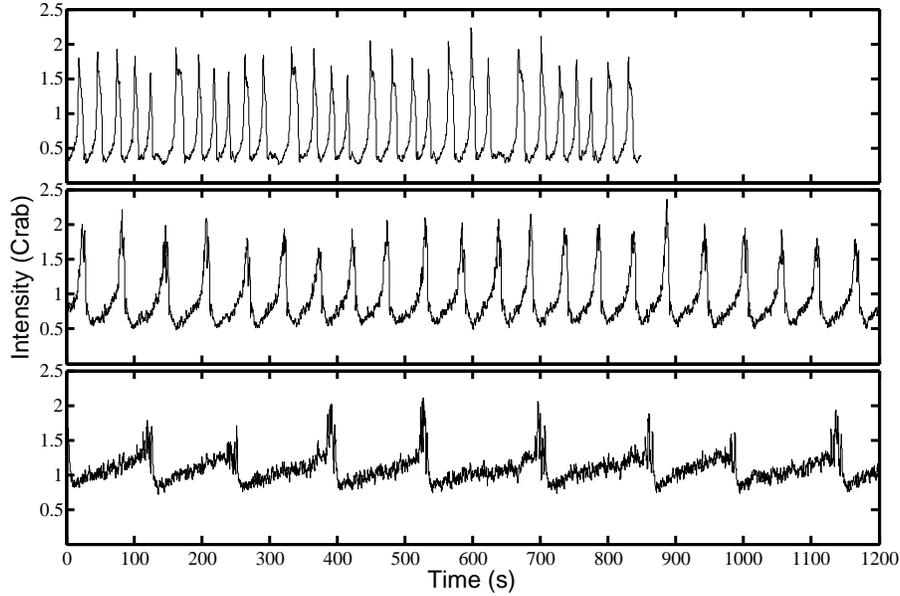}
\caption{Examples of light curves for different recurrent timescales. The data
come from observations 92702-01-14-00 ($R_{\rm in} = 63.8_{-2.3}^{+2.4}$ km,
upper panel), 95701-01-03-00 ($R_{\rm in} = 87.5_{-2.7}^{+2.1}$ km, middle
panel), and 91701-01-52-01 ($R_{\rm in} = 91.5_{-2.7}^{+2.7}$ km, bottom
panel), respectively. \label{lc}}
\end{figure*}

Examples of light curves with different recurrent time scales are displayed in
Figure \ref{lc}, in which the correlations described above are obvious. It can
be also found that the change of the recurrent time is mainly because of the
variation of rising time scale, namely from the $C_{\rm min}$ increasing to the
$C_{\rm max}$. In contrast, the decaying time scales (i.e. from $C_{\rm max}$
to $C_{\rm min}$) are similar for different values of $T_{\rm rec}$.

\subsection{Spectral parameters}

\begin{figure*}
\centering
\includegraphics[width=5.5cm]{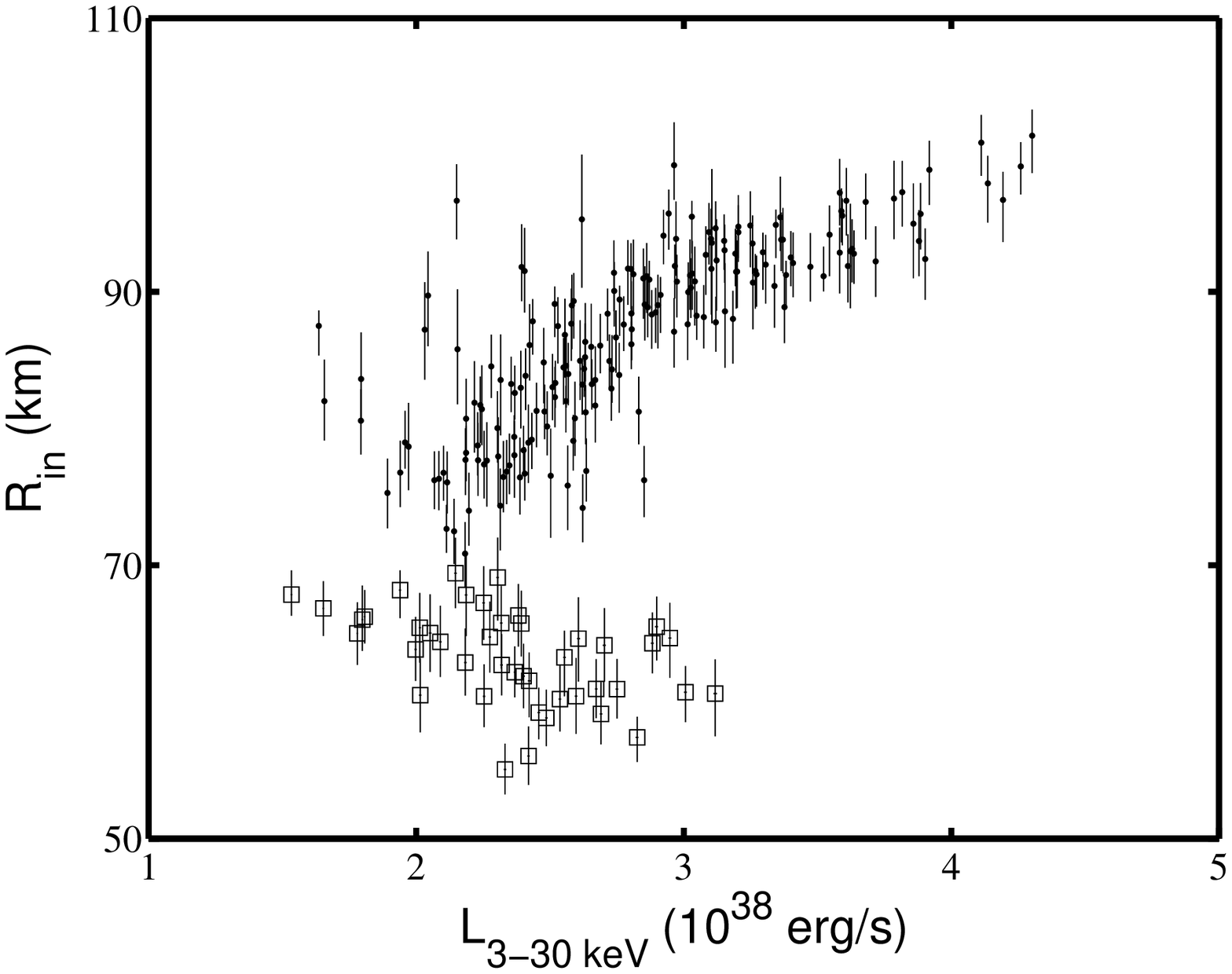}
\includegraphics[width=5.5cm]{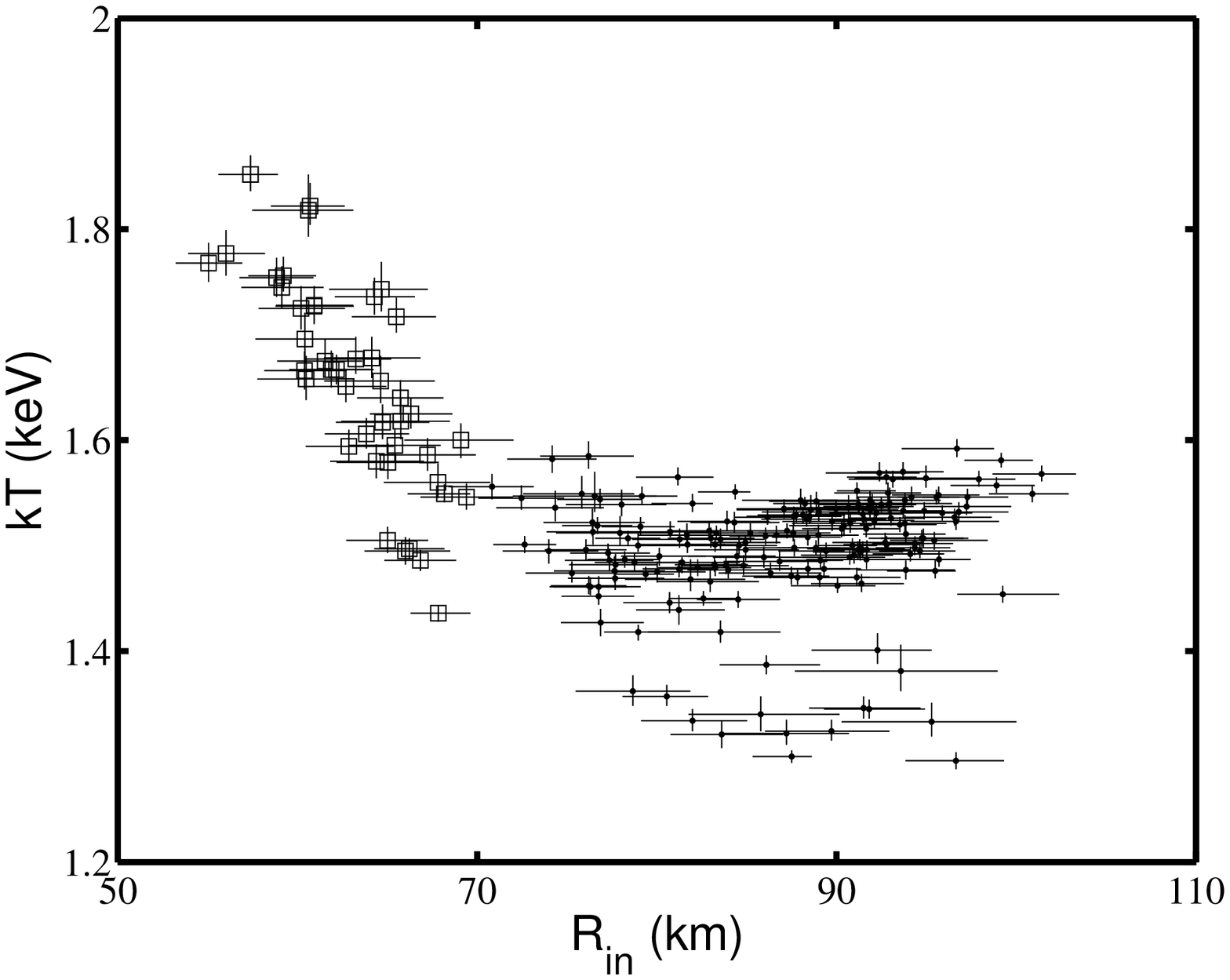}
\includegraphics[width=5.5cm]{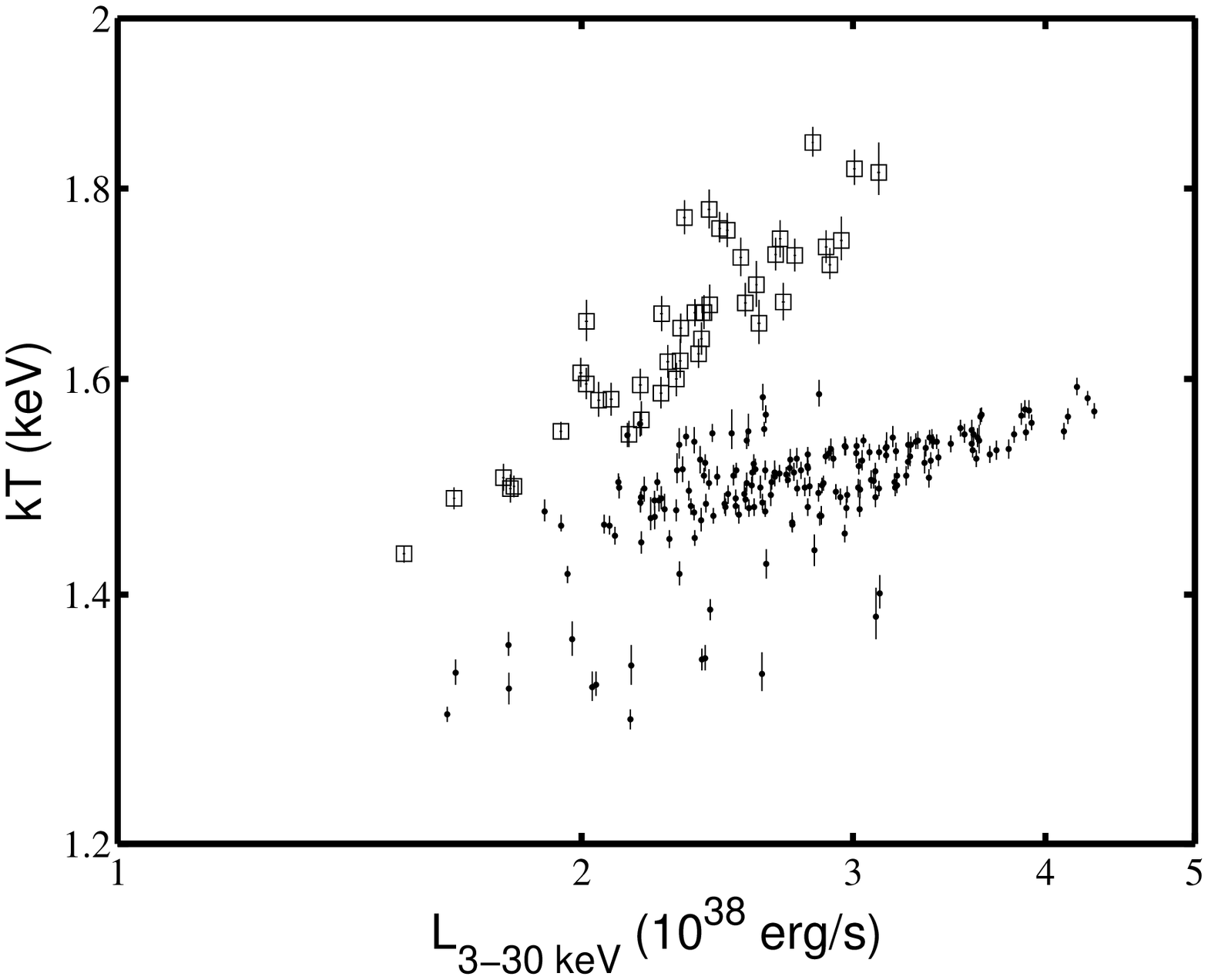}
\includegraphics[width=5.5cm]{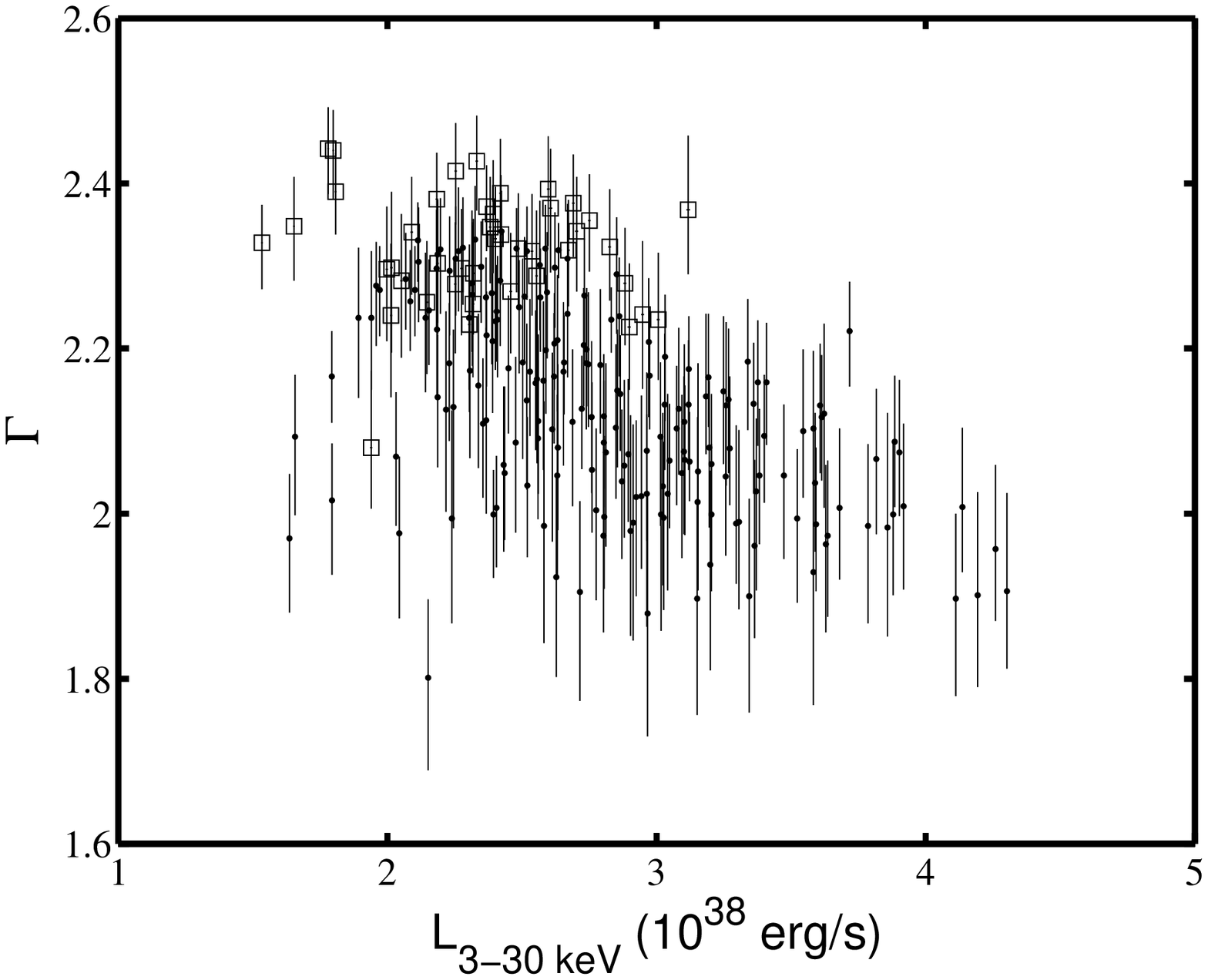}
\includegraphics[width=5.5cm]{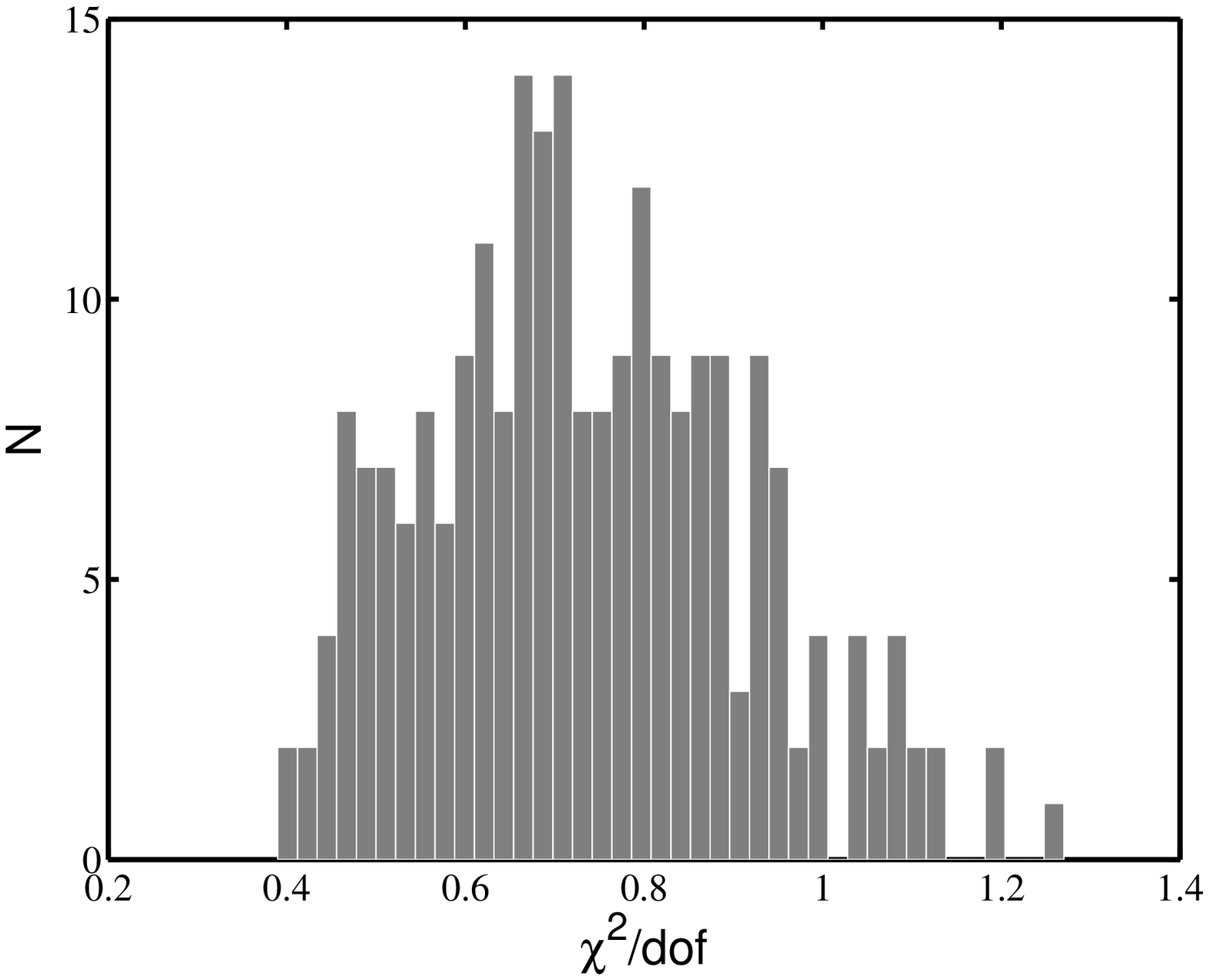}
\caption{Correlations between the spectral parameters and distribution
of the reduced $\chi^{2}$ with the fitting model of {\it diskbb+powerlaw}
(right bottom panel). Open squares and filled cycles correspond to the data
with $R_{\rm in}$ being smaller and larger than 70 km, respectively. All
parameters are plotted with their 1$\sigma$ errors. \label{spec_corr}}
\end{figure*}

The spectral fitting results are presented in Figure \ref{spec_corr}. The
unabsorbed power-law luminosity ($L_{\rm PL}$) and the disk luminosity increase
simultaneously with the total luminosity ($L_{\rm 3-30~keV}$). The apparent
inner disk radius ($R_{\rm in}$) is related to the disk normalization as
$N_{\rm disk} = \frac{1}{f^{4}}(\frac{R_{\rm in}}{D_{10}})^{2} \cos \theta$,
where $D_{10}$ is the distance in units of 10 kpc. Here we adopt an the
inclination angle $\theta = 66\degr$ \citep{Fender99}, a color correction
factor $f = 1.9$ \citep{Neilsen11}, and a distance of 8.6 kpc \citep{Reid14} in
the calculation. There are two branches shown in the diagrams of $L_{\rm
3-30~keV} - R_{\rm in}$ and $L_{\rm 3-30~keV} - kT$: (1) The inner disk remains
at a small radius ($R_{\rm in} \leq 70$ km, hereafter S branch), $kT$ increases
with the X-ray luminosity, and the X-ray emissions are fainter and softer
(larger $\Gamma$). (2) The majority of data (189 out of 233) draw the other
branch (L branch), the apparent inner edge of disk recedes, and the temperature
increases slowly at the high luminosity. Meanwhile, the photon index ($\Gamma$)
of nonthermal component is anticorrelated with the X-ray luminosity.

Intriguingly, there are the positive correlations between $T_{\rm rec}$,
$R_{\rm in}$ and the X-ray luminosity as well. Additionally, $T_{\rm rec}$ has
an even tighter correlation with the power-law luminosity. Since the
uncertainty on $T_{\rm rec}$ is much larger than on $L_{\rm 3-30~keV}$, we
carry out a power-law fit to the $T_{\rm rec} \propto L_{\rm 3-30~keV}^{n}$
relation by taking the error on $T_{\rm rec}$ into account, yielding an index
of $n = 1.10\pm0.10$. Meanwhile, $T_{\rm rec}$ is proportional to $L_{\rm
PL}^{1.66\pm0.09}$.

\begin{figure*}
\centering
\includegraphics[width=5.5cm]{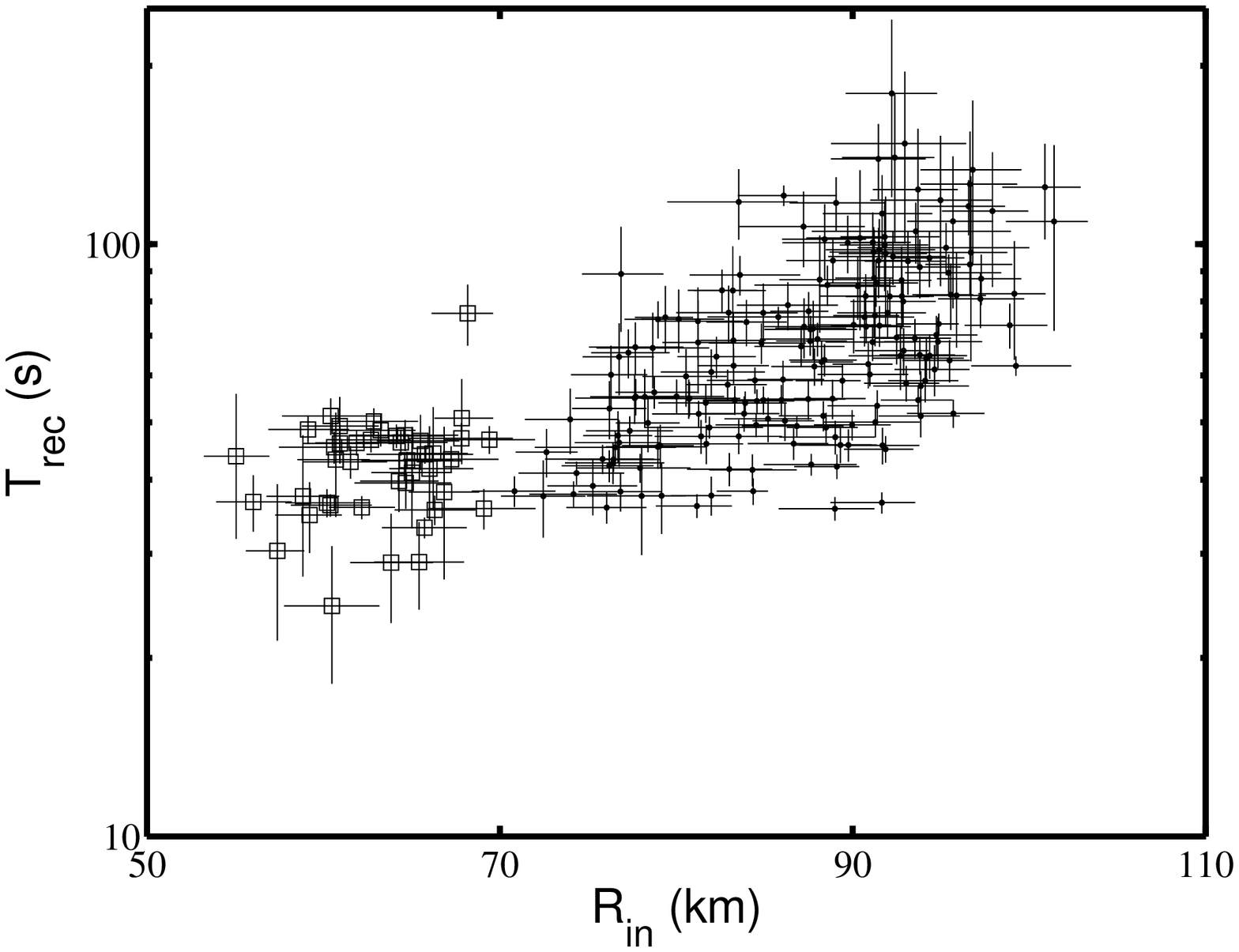}
\includegraphics[width=5.5cm]{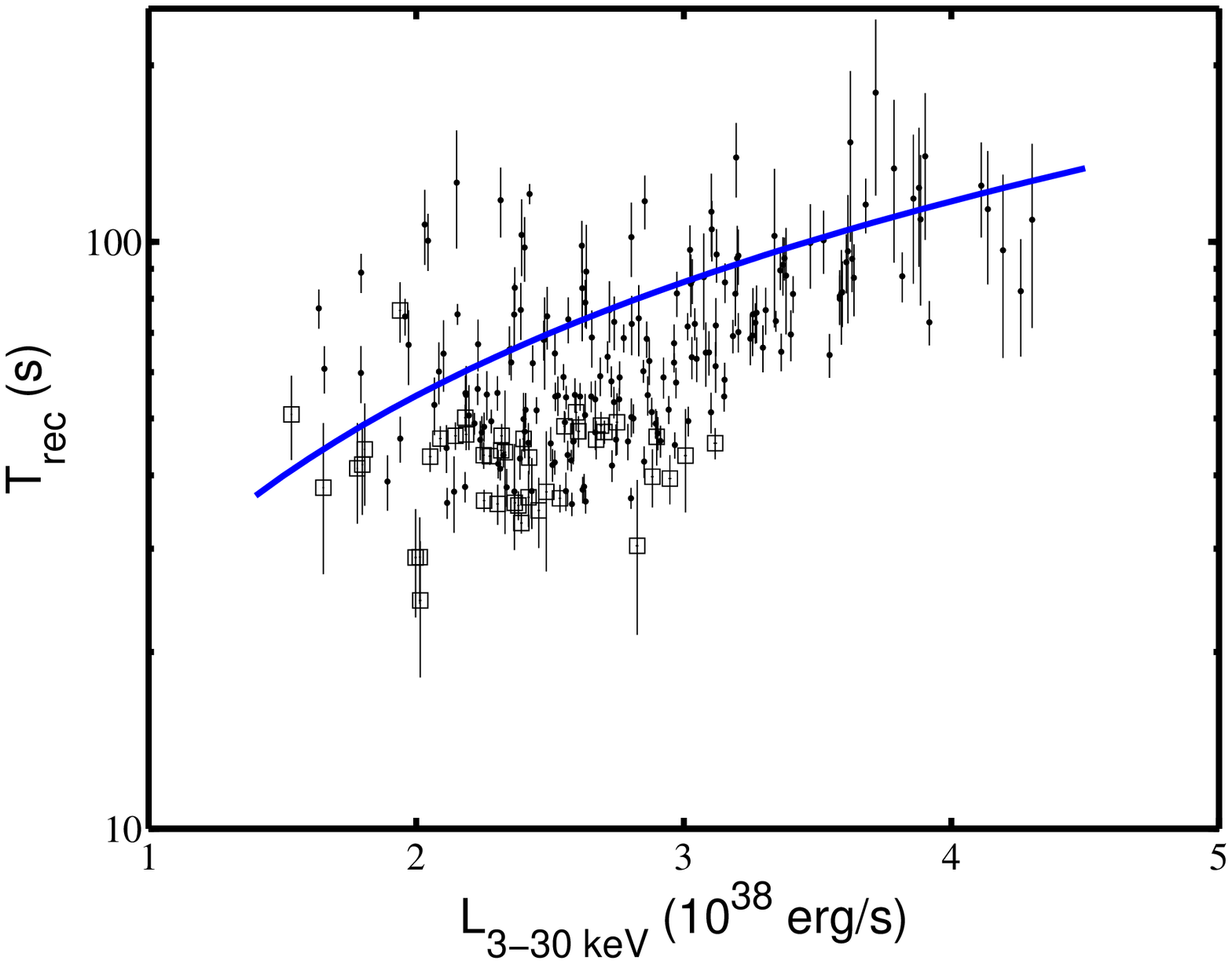}
\includegraphics[width=5.5cm]{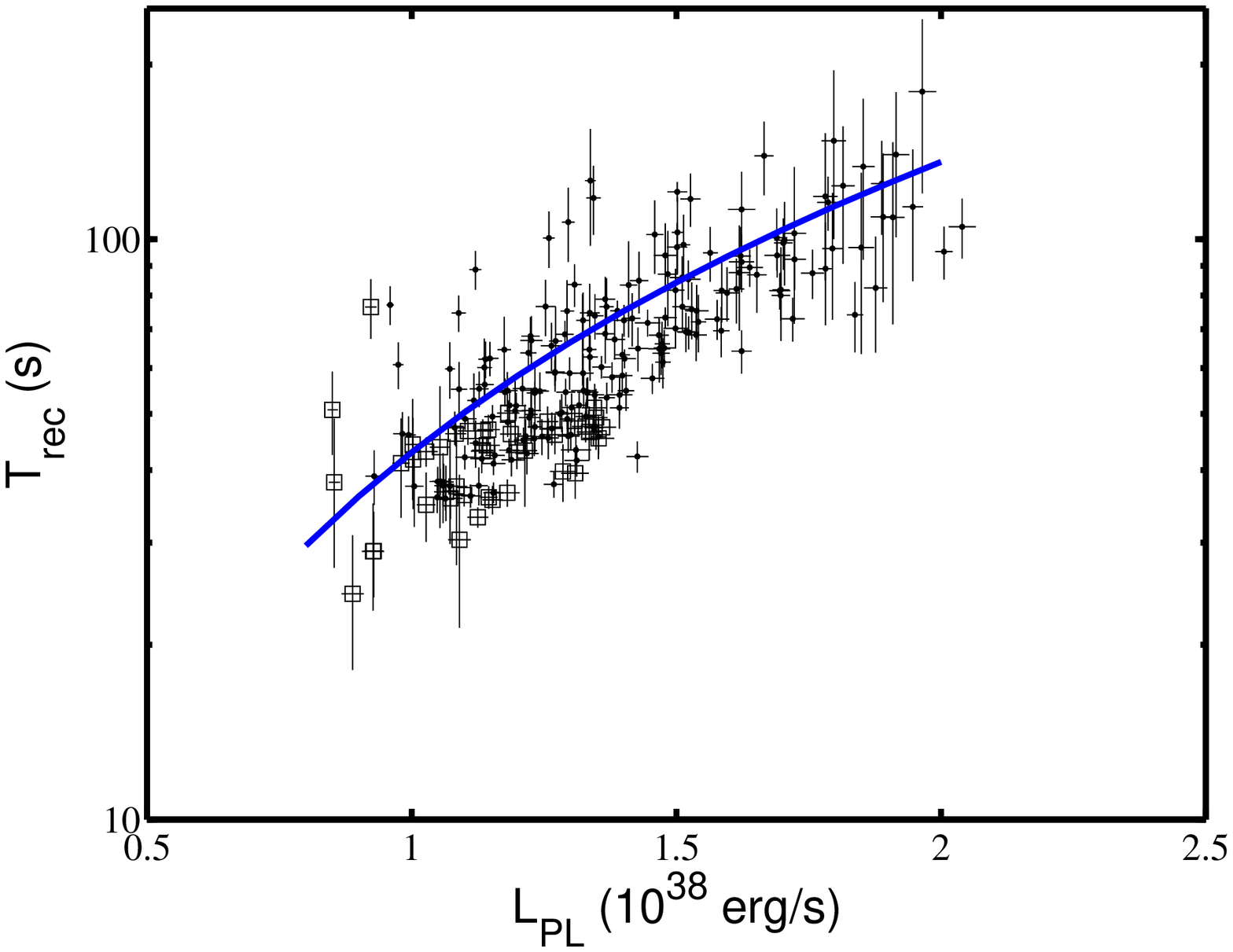}
\caption{Recurrence time increases with $R_{\rm in}$, $L_{\rm 3-30~keV}$, and
$L_{\rm PL}$. The power-law function is employed to fit the $L_{3-30
{\rm keV}} - T_{\rm rec}$ and $L_{\rm PL} - T_{\rm rec}$ correlations (blue
lines). \label{heartbeat_corr}}
\end{figure*}

\begin{figure*}
\centering
\includegraphics[width=5.5cm]{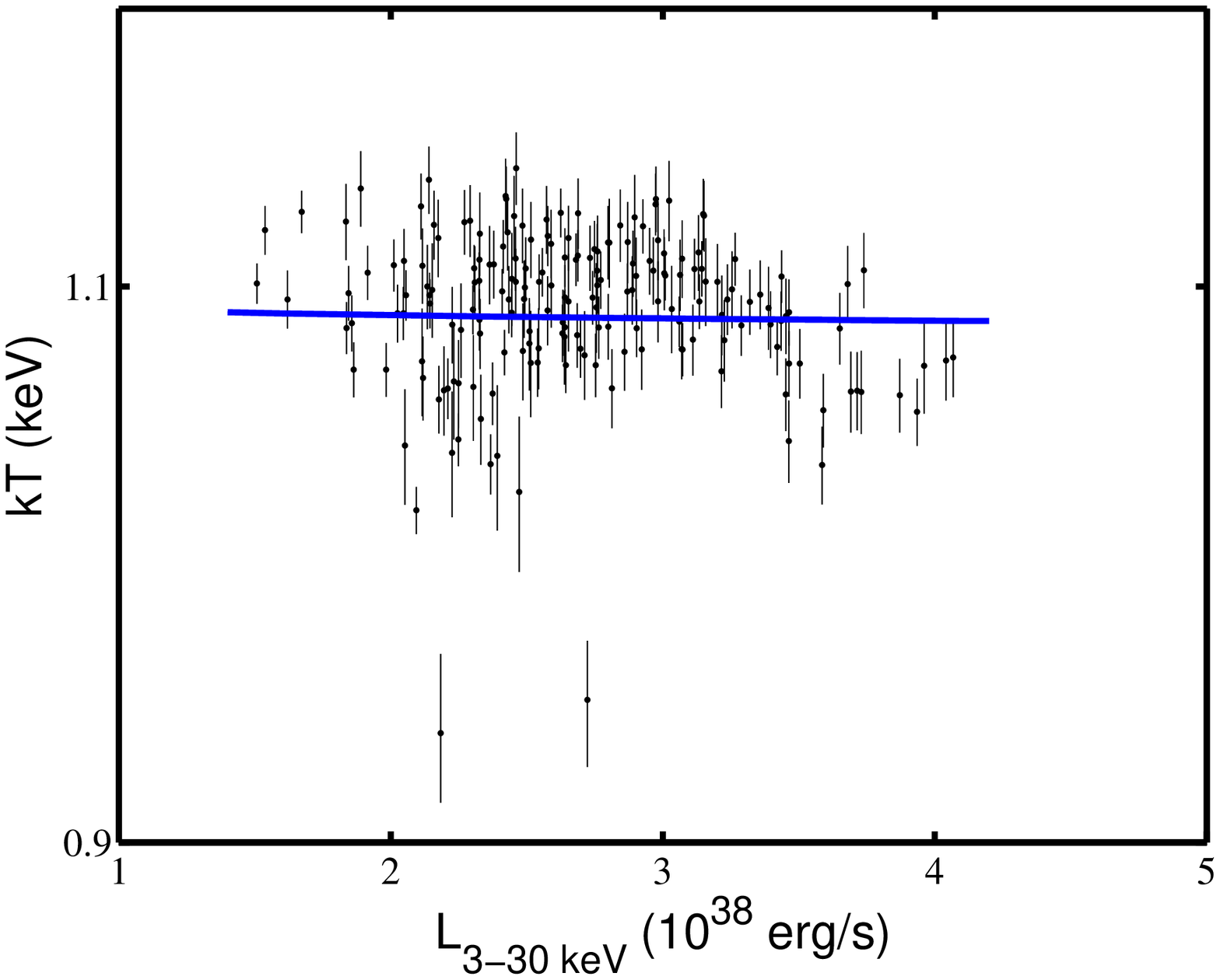}
\includegraphics[width=5.5cm]{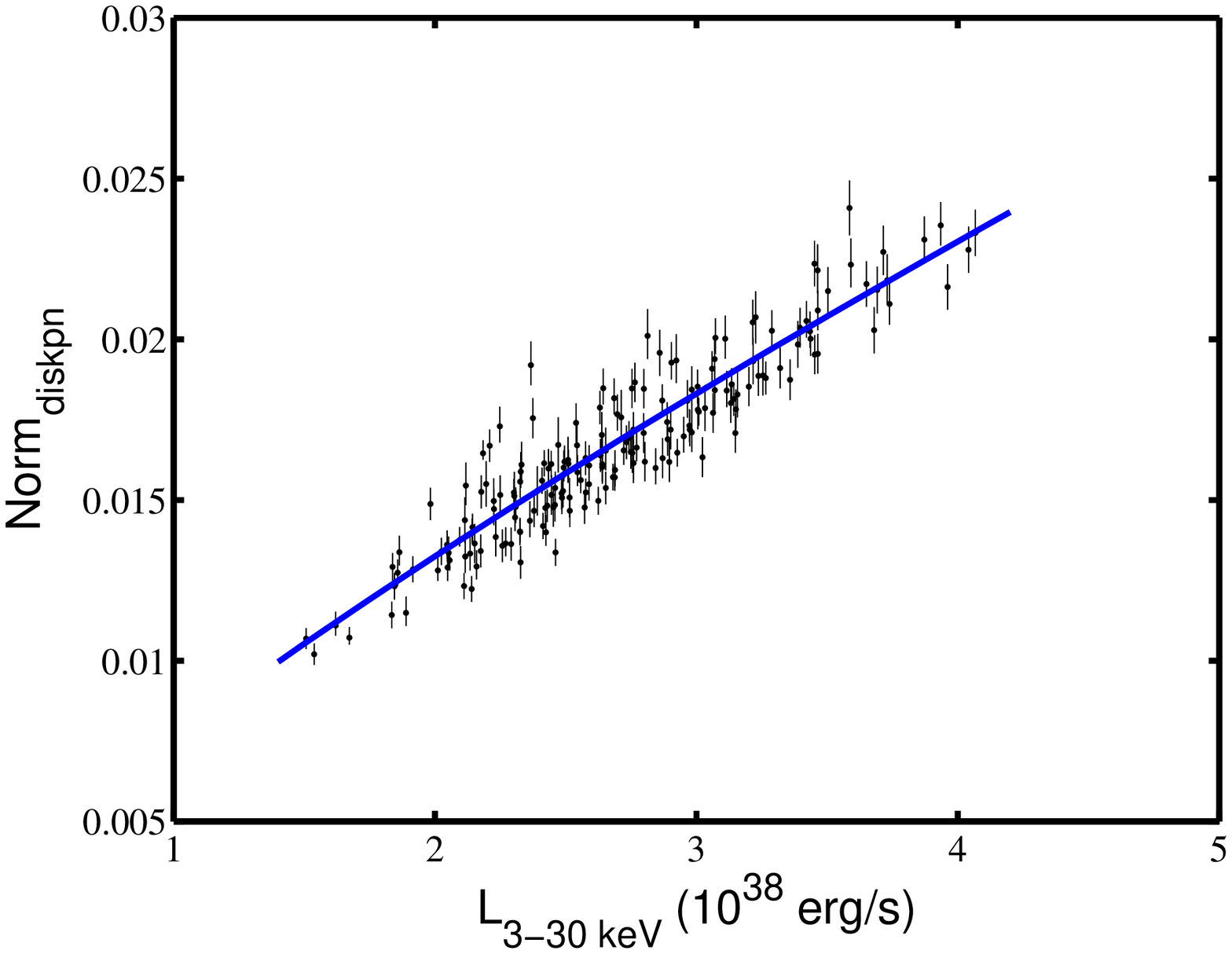}
\includegraphics[width=5.5cm]{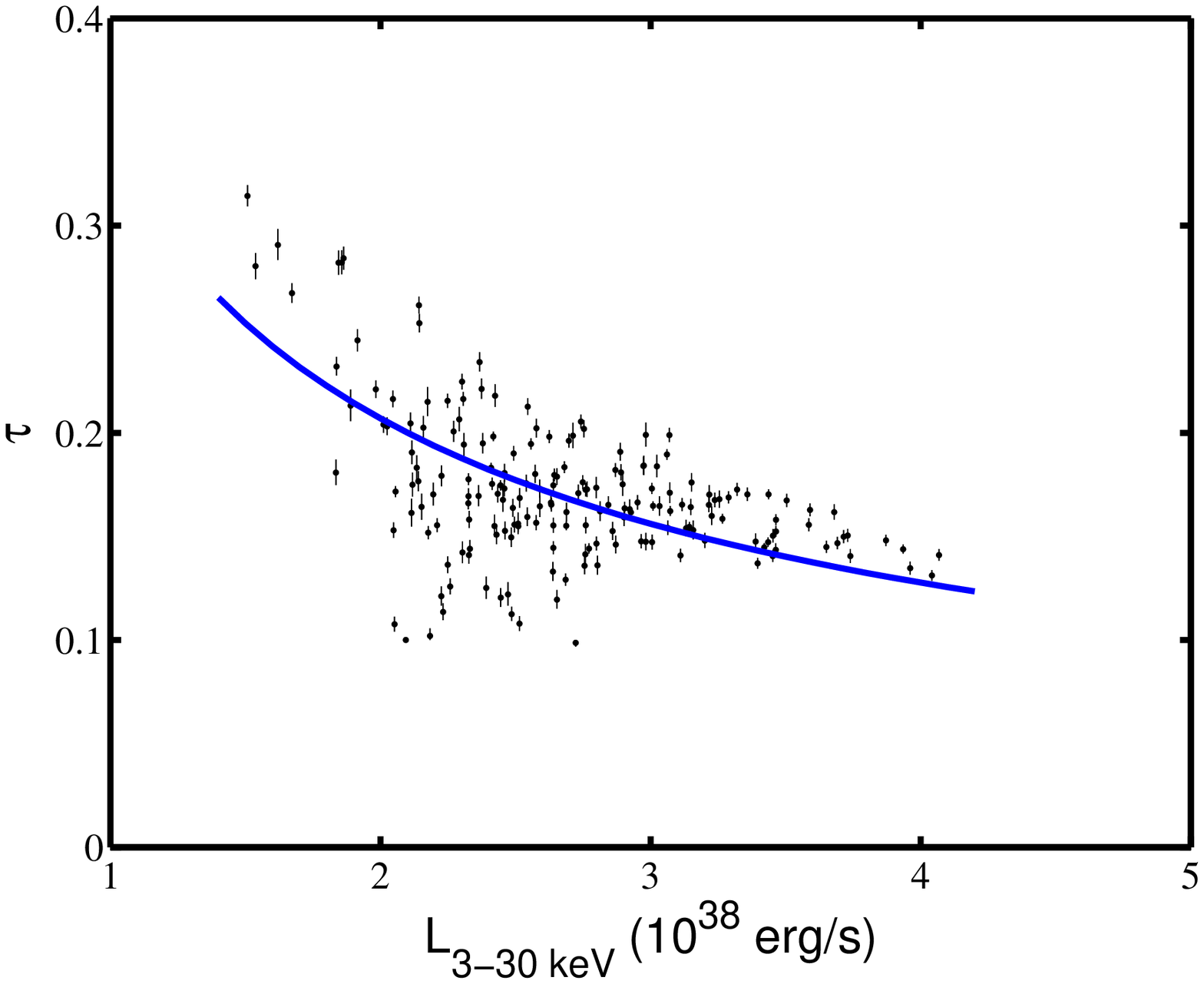}
\includegraphics[width=5.5cm]{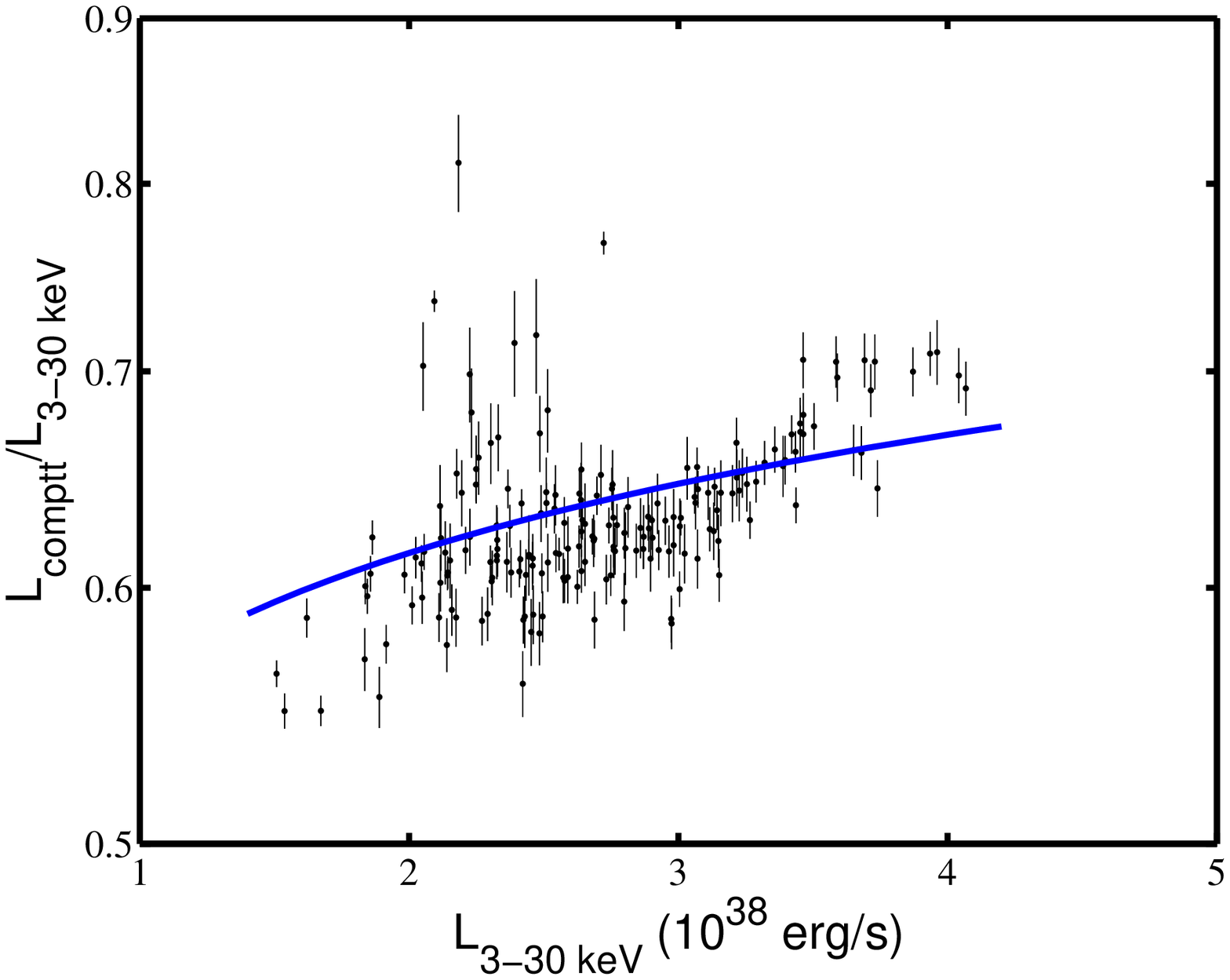}
\includegraphics[width=5.5cm]{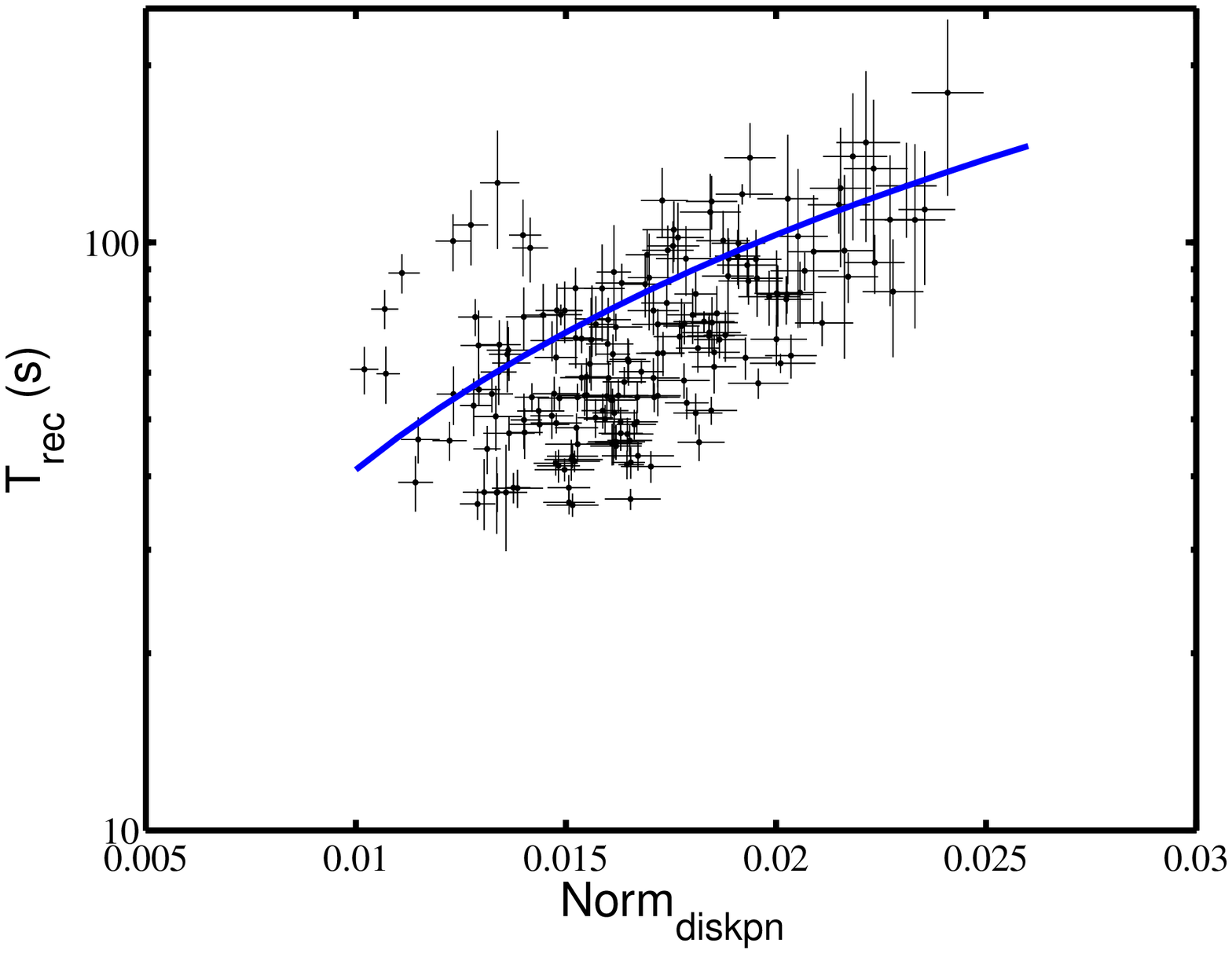}
\includegraphics[width=5.5cm]{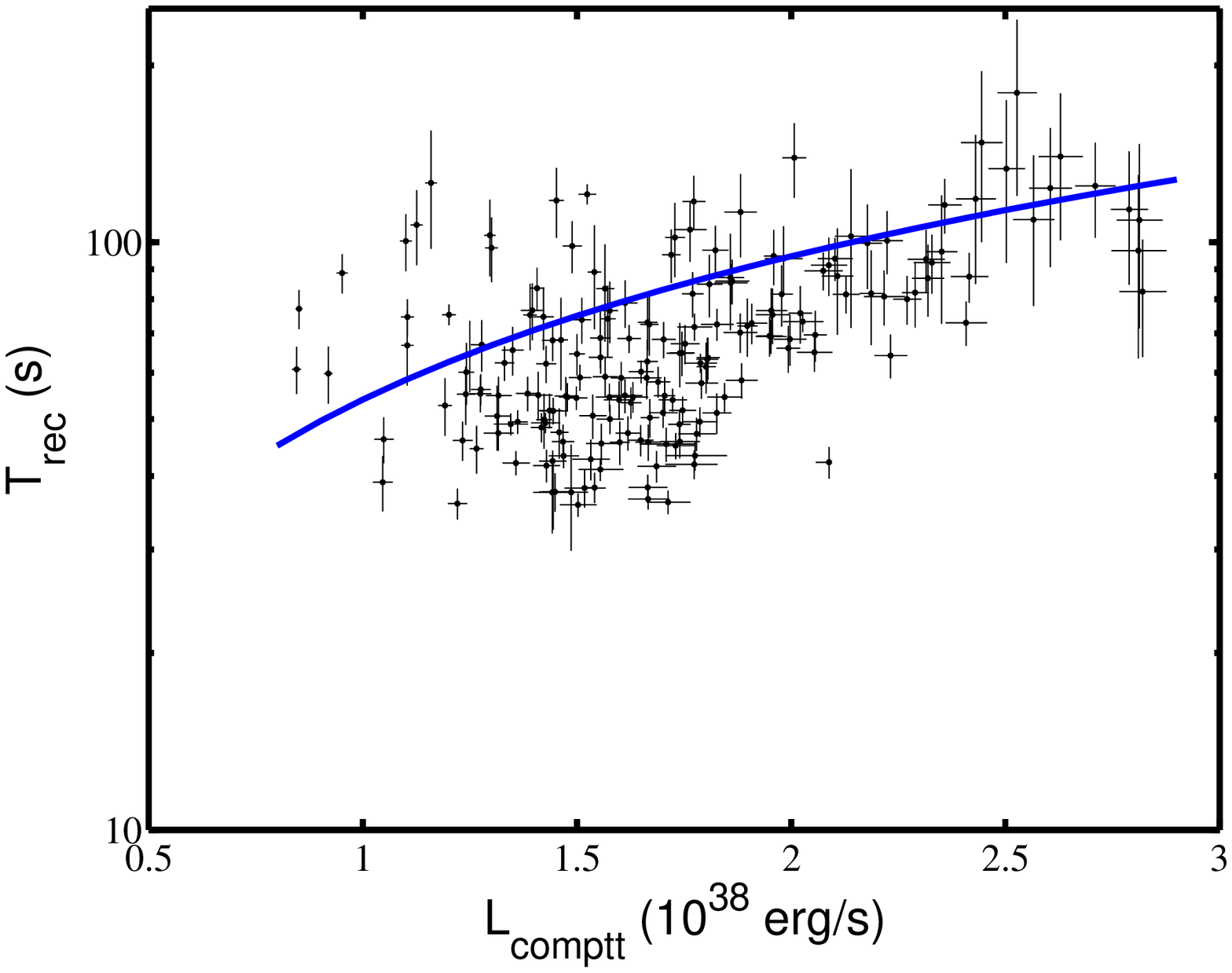}
\caption{Correlations between the spectral parameters of {\it diskpn+comptt}
model and the structure parameters of light curves. Blue lines correspond to
the power-law fittings: $kT \propto L_{\rm 3-30 keV}^{0.00\pm0.01}$, $Norm_{\rm
diskpn} \propto L_{\rm 3-30 keV}^{0.80\pm0.03}$, $\tau \propto L_{\rm 3-30
keV}^{-0.70\pm0.06}$, $L_{\rm comptt}/L_{\rm 3-30 keV} \propto L_{\rm 3-30
keV}^{0.12\pm0.03}$,  $T_{\rm rec} \propto Norm_{\rm diskpn}^{1.38\pm0.11}$,
and $T_{\rm rec} \propto L_{\rm comptt}^{0.81\pm0.08}$. \label{comptt}}
\end{figure*}

It is worth noting that the change of apparent inner disk radius can be
interpreted as a varying color correction factor without moving the intrinsic
inner disk radius \citep[e.g. ][]{Merloni00, Zoghbi16}. In order to check this
scenario, we model the disk component with {\it diskpn} \citep{Gierlinski99},
in which the disk normalization is linked to the color correction factor:
$Norm_{\rm diskpn} = \frac{M^{2} \cos \theta}{D^{2} f^{4}}$. The inner disk
radius is fixed to the bottom limit $6~R_{\rm g}$, although it is much smaller
due to high spinning rate \citep{McClintock06}. In addition, the power-law
model is replaced by a physical Comptonization model ({\it
tbabs*(diskpn+comptt)} in XSPEC). We tie the seed temperature ($kT_{\rm seed}$)
to the disk temperature and fix the plasma temperature to 50 keV because it
cannot be constrained by our data. The data in the S branch are relatively
soft, and the model described above cannot provide an adequate fit to them
unless the $kT_{\rm seed}$ varies freely and becomes hotter than the disk
temperature; however, it is inconsistent. Henceforth, we only fit the data in
the L branch. With the X-ray luminosity increasing, the optical depth
decreases, while the contribution of the Comptonized component and the
$Norm_{\rm diskpn}$  increase (Figure \ref{comptt}). That is, the color
correction factor decreases with the X-ray luminosity. Further discussion is
presented in the next section.

\section{Discussion and Conclusion}

\subsection{Movable accretion disk?}

Plenty of works have been carried out to study the heartbeat state by
investigating the individual observation \citep[e.g. ][]{Massaro10, Neilsen11,
Neilsen12, Zoghbi16, Yan17, Yan18}. The regular flares are generally attributed
to the thermal-viscous instability \citep[i.e. the Lightman-Eardley
instability; ][]{Lightman74} when the inner region is dominated by radiation
pressure at high luminosities \citep[e.g. ][]{Belloni97, Janiuk05,
Grzedzielski17}. Phenomenologically, the burst sequence could be interpreted as
the rapid emptying of the inner portion of the accretion disk, followed by a
slower refilling of the inner region, and it finishes a cycle on the viscous
time scale \citep{Belloni97}. Theoretically, some groups successfully
reproduced the oscillating light curve with the proper cycle and the burst
amplitude by modifying the classical description of $\alpha$ viscosity
\citep[e.g. ][]{Honma91, Szuszkiewicz98, Merloni06, Zheng11, Grzedzielski17}.
However, such models cannot account for the growth of $R_{\rm in}$ at the slow
rise stage as suggested by the phase-resolved study \citep{Neilsen11}.
Moreover, our statistical analysis finds the anticorrelation between $T_{\rm
rec}$ and the amplitude (Figure \ref{lc_corr}), which is in contrast with the
prediction of the modified viscosity model \citep{Grzedzielski17}.

Besides the radiation pressure instability, \cite{Neilsen11} argued that a
local Eddington effect \citep{Fukue04, Heinzeller07} is required to explain
some X-ray properties of the heartbeat state. When the luminosity is larger
than the critical value ($\sim 0.3-1~ L_{\rm Edd}$), the radiation pressure
could either push the disk outward or generate the optically thick outflows
\citep[e.g. ][]{Poutanen07, Weng14, Urquhart16}, or the disk might become thick
and block the inner region \citep[e.g. ][]{McClintock06, Gu12}. As a result,
the apparent size of the thermal component increases with the luminosity, and
the X-ray emissions become harder in the meantime \citep{Soria07, Middleton15}.
In the S branch, the X-ray luminosity is below and close to the critical
luminosity, softer emissions and a smaller apparent accretion disk are
observed. If the recurrence time of heartbeat flare is related to the viscous
time of inner disk radius, we would expect that $T_{\rm rec}$ increases with
increasing $R_{\rm in}$ and the X-ray luminosity as shown in Figure
\ref{heartbeat_corr}.

\subsection{Variable corona?}

The advection, coronal dissipation, and outflows play important roles in the
stabilization of accretion flows at the high luminosity state \citep[e.g.
][]{Janiuk02}. Investigating the quasi-simultaneous radio and X-ray data,
\cite{Vadawale03} achieved the association between jet and corona. For
instance, the central Compton cloud (corona) is ejected during the soft X-ray
dips, which are preceded by a radio-loud hard state. They argued that the
change in the corona can account for the X-ray variations that were previously
attributed to the accretion disk. Such a scenario, in particular, is supported
by the absence of both the Comptionized component and low-frequency QPOs during
the X-ray dips. Note that low-frequency QPO signals detected in low-mass X-ray
binaries are generally connected with the hard component but not (or sometimes
indirectly linked with) the thermal accretion disk \citep[e.g. ][]{Belloni14,
Motta15, Zhao16, Zhang17, Yan18}.

Our results indicate that, for observations with different $T_{\rm rec}$, the
main difference is from the slow rise stage, while the source takes almost the
same time to return from the peak luminosity to the minimum (Figure \ref{lc}).
The change of $C_{\rm min}$ would further point to the variation of corona and
jet properties for different $T_{\rm rec}$ oscillations. The tighter
correlation between $R_{\rm in}$ and the power-law luminosity (Figure
\ref{heartbeat_corr}) also supports that the limit cycle is driven by the
nonthermal emissions rather than the thermal component. Meanwhile, $C_{\rm
max}$ remains nearly constant, indicating that the bolometric luminosity of
GRS~1915+105 reaches its Eddington luminosity.

The variation of the corona does not only affect the Comptionized component
itself, but also modifies the property of the observed disk thermal emission
via Compton scattering \citep[e.g. ][]{Shimura95, Nayakshin00}. The spectral
hardening factor $f$ is by no means constant, as we assumed in the $R_{\rm in}$
calculation. It should depend on the accretion rate, the fraction of the
accreted power released in the corona, the corona reflecting, etc \citep[e.g.
][]{Merloni00, Davis05}. When the actual inner boundary of the accretion disk
is fixed, the color disk radius could vary by more than a factor of 4 with
different accretion parameters \citep{Merloni00}.

Additionally, \cite{Zoghbi16} argued that a change in composition of disk
atmosphere could make an even larger change in the color correction factor. As
the disk temperature and flux increase, more iron ions are elevated up by the
radiation force, increasing the opacity in the upper disk and resulting in
smaller $f$. They further suggested that the evolution of $R_{\rm in}$ inferred
from the disk blackbody component was artificial, while the actual inner edge
was kept at a small radius ($\sim 1.1~R_{\rm g}$) in the heartbeat state
according to the reflection measurement. Specifically, \cite{Zoghbi16} proposed
that the inner disk radius remained constant and the corona changed size during
oscillations. If the corona is smaller and closer to the central BH, more
photons are dragged by the strong gravity and hit the disk; therefore, fewer
photons can reach the observer. This model is in qualitative agreement with our
results presented in Figure \ref{comptt}, i.e. the {\it comptt}-to-total-flux
ratio and $T_{\rm rec}$ increase with the total X-ray luminosity. The change of
apparent disk size shown in Figure \ref{spec_corr} can be interpreted as the
variation of the color correction factor. The correlation between the
X-ray/disk flux and $Norm_{\rm diskpn}$ (Figure \ref{comptt}) indicates that
the factor $f$ decreases from $\sim 3.0$ to $\sim 2.4$ with the increasing
luminosity. But note that if the inner disk radius was fixed to 6 $R_{\rm g}$,
the actual value of $f$ would be smaller with smaller inner disk radius.

Although the varying corona scenario offers a promising explanation for the
results given by our statistical analysis, we cannot completely rule out the
evolving disk model. In addition, a number of questions remain to be answered.
(1) What is the origin of variations in the corona (and the disk)? (2) If the
low-frequency QPOs and $T_{\rm rec}$ are account for different time scales of
corona, e.g. dynamical time and thermal/viscous time; but their correlation is
unknown at the current stage. (3) It has been pointed out that the reflection
component is prominent in the heartbeat state \citep{Zoghbi16}. The evolution
trend of the disk component and the total X-ray luminosity are insensitive to
models with or without the reflection. However, the parameters of the thermal
and Comptonization components might change when the reflection is included or
excluded in fitting models. Unfortunately, we cannot study the reflection
spectrum with the {\it RXTE}/PCA data due to its low energy resolution. On the
other hand, the single {\it Nustar} observation is inadequate to explore the
spectral/temporal evolution with different $T_{\rm rec}$. Thus, we would like
to suggest that multiple high quality data (high-energy/time resolution) from
Insight-{\it HXMT} and {\it Nustar} are required to probe the peculiar state
deeper.

\acknowledgements {We thank the referee for helpful comments and suggestions.
We thank Drs. Shu-Ping Yan, Ming-Yu Ge, and You-Li Tuo for many valuable
suggestions. This work is supported by the National Natural Science Foundation
of China under grants 11673013, 11703014, 11433005, 11573023, and 11873032, and
the Natural Science Foundation from Jiangsu Province of China (Grant No.
BK20171028).}

{\it Facilities: RXTE}

{\it Software:} \textsc{heasoft}.

\end{document}